\documentclass[aps,showpacs,twocolumn,superscriptaddress,floatfix]{revtex4}
\usepackage{graphicx}
\usepackage{amsmath,amssymb}
\usepackage[utf8]{inputenc}
\renewcommand\vec[1]{\mathbf{#1}}
\newcommand\tens[1]{\mathsf{#1}}
\begin{document}
\title{Generic flow profiles induced by a beating cilium}
\author{Andrej Vilfan}
\email{andrej.vilfan@ijs.si}         
\affiliation{J. Stefan Institute, Jamova 39, 1000 Ljubljana, Slovenia}
\affiliation{Faculty of Mathematics and Physics, University of Ljubljana, Jadranska 29,
  1000 Ljubljana, Slovenia}
\begin{abstract}We describe a multipole expansion for the low Reynolds number
  fluid flows generated by a localized source embedded in a plane with a
  no-slip boundary condition. It contains 3 independent terms that fall
  quadratically with the distance and 6 terms that fall with the third
  power. Within this framework we discuss the flows induced by a beating
  cilium described in different ways: a small particle circling on an
  elliptical trajectory, a thin rod and a general ciliary beating pattern. We
  identify the flow modes present based on the symmetry properties of the
  ciliary beat.
\end{abstract}
 % end of PACS codes
\maketitle

\section{Introduction}
\label{intro}
Cilia are thin cellular protrusions that beat in an asymmetric periodic
fashion in order to propel the surrounding fluid
\cite{Gray1928,Brennen.Winet1977,Sleigh74}. They are involved in swimming and
feeding of a number of protozoa and also have many crucial functions in
vertebrates. These include the mucous clearance in
respiratory pathways, transport of an egg cell in Fallopian tubes, left-right
symmetry breaking in embryonic development
\cite{afzelius99,Hirokawa.Takeda2006,Supatto.Vermot2008}, and, recently
discovered, otolith formation in hearing organs
\cite{Colantonio.Hill2009}. Cilia have inspired designs for microfluidic pumps
and mixers using similar beating patterns
\cite{Vilfan.Babic2010,Kokot.Babic2011,Gauger.Stark2009,Toonder.Anderson2008,Coq.Bartolo2011,Hussong.Westerweel2011,Shields.Superfine2010}.

The beating pattern of a single cilium can be very complex and generally
consists of a working stroke during which the stretched cilium moves the fluid
in the direction of pumping and a recovery stroke during which the cilium
folds and returns to the origin by sweeping along the surface, thereby
minimizing the drag as well as backward flow. Especially when working in large
ensembles, the beating patterns found in nature are remarkably close to the
theoretically calculated optimum \cite{Osterman.Vilfan2011}.  Despite this
complexity, the flows can be described with a small number of generic terms
in the far field, i.e., when we observe the flow at a distance $r$ which is
sufficiently larger than the cilium length $L$. In the leading order the flow
velocity falls as ${\cal O}(r^{-2})$ \cite{vilfan2006a}. 

In this paper we go beyond the quadratic order and provide a complete set of
functions that describe the flow in the presence of a no-slip boundary. We
then specifically discuss the ${\cal O}(r^{-3})$ terms that are present with
different beating patterns, depending on their symmetry properties.

\section{Multipole expansion}
\subsection{In unbounded fluid}

In the low Reynolds number regime, the motion of an incompressible fluid is
described by the Stokes equation and the incompressibility condition:
\begin{align}
\eta \Delta \vec{v} &=\vec{\nabla} p\label{eq:stokes}\\
\vec{\nabla} \cdot \vec{v}&=0\;.
\end{align}
To derive the general form of the flow induced by a localized distribution of
forces and/or sources we follow the approach of Lamb \cite{Lamb}, also used by
Happel and Brenner \cite{Happel.Brenner}. Note that this is just one of many
complete sets of far-field solutions (see Ref. \cite{Pozrikidis1992} for a
derivation in Cartesian coordinates). Although the basis set we will derive
can be obtained more directly as spatial derivatives of known solutions
(Stokeslet and source), the following construction has several
advantages. First, we know from the beginning on that we are dealing with a
complete set of linearly independent solutions. Second, these solutions will
appear classified by their angular symmetry, which are useful for the
description of flows induced by cilia which also share some of these
symmetries. And finally, the construction allows us to distinguish between
different solutions (e.g., with or without a pressure gradient).

Multiplying Eq.~(\ref{eq:stokes}) with
$\vec{\nabla}\cdot$ immediately leads to a Laplace equation for the pressure
\begin{equation}
  \label{eq:pressurelaplace}
  \Delta p=0\;.
\end{equation}
The solution of the homogeneous equation in the absence of pressure gradients
($\Delta \vec{v}=0$) can be written as $\vec{v}=\vec{\nabla}\times(\vec{r}
\chi) +\vec{\nabla} \Phi$, where $\chi$ and $\Phi$ are two solutions of the
Laplace equation. In total we need three sets of harmonic functions ($p$,
$\chi$ and $\Phi$) to construct the general solution of the Stokes equation in
unbounded space.  We now introduce a spherical coordinate system such that
\begin{equation}
  \label{eq:coords}
  \vec{x} = \left( \begin{array}{c} r \sin \theta \cos \phi\\
      r \sin \theta \sin \phi \\
      r \cos \theta
    \end{array}\right)\;.
  \end{equation}
  If there is no external flow, i.e, the $\lim_{r\to \infty} \vec{v}(\vec{x})
  =0$, the multipole expansion of the velocity field reads
\begin{equation}
  \label{eq:expansion}
  \vec{v}=\sum_{l=0}^{\infty} \vec{v}_l\qquad \text{with}\; \vec{v}_l \sim r^{-(l+1)}
\end{equation}
with
\begin{multline}
\label{eq:v}
   \vec{v}_l= \nabla \times (\vec{r} \chi_{l}) + \nabla
   \Phi_{l-1} \\+
   \frac{1-l}{2\eta (1+l) (1+2l)} r^2 \nabla p_{l+1} +
   \frac{2+l}{\eta (1+l) (1+2l)} \vec{r} p_{l+1} 
 \end{multline}
 and
\begin{align}
  \chi_{l}&=\frac 1 { r^{l+1}} \sum_{m=-l}^l  A_{lm} e^{i m \phi} P_l^m (\cos \theta)\\
  \Phi_{l-1}&=\frac 1 { r^{l}} \sum_{m=-(l-1)}^{(l-1)}  B_{lm}   e^{i m \phi} P_{l-1}^m
  (\cos \theta)\\
  p_{l+1}&=\eta (1+l)(1+2l) \frac 1 { r^{l+2}} \sum_{m=-(l+1)}^{(l+1)}  C_{lm}   e^{i m \phi} P_{l+1}^m
  (\cos \theta)\;.
\end{align}
The coefficients $A_{lm}$ are defined for $l\ge 1$ and $|m|\le l$, $B_{lm}$
are defined for $l\ge 1$ and $|m|\le l-1$ and $C_{lm}$ for $l\ge 0$ and
$|m|\le l+1$.  Using an elementary vector identity the first term in
(\ref{eq:v}) can alternatively be written as $(\vec{\nabla} \chi_l)\times
\vec{r}$.  Inserting these terms into Eq.~(\ref{eq:v}) gives the expression
for the velocity
\begin{multline}
  \vec{v}_{lm}=\frac{e^{i m \phi}}{r^{l+1}} \Bigl[\Bigl( A_{lm} P_l^{\prime m}
  (\cos\theta ) \sin\theta +B_{lm} \frac{im}{\cos \theta} P_{l-1}^{m}(\cos
  \theta) \\+ C_{lm}\frac{1-l}{2} \frac{im}{\sin\theta} P_{l+1}^{m}(\cos
  \theta) \Bigr)
  \hat e_\phi \\
  + \Bigl( A_{lm} \frac{im}{\sin\theta} P_l^m (\cos\theta ) -B_{lm}
  P_{l-1}^{\prime m}(\cos \theta) \sin\theta \\ - C_{lm}\frac{1-l}{2}
  P_{l+1}^{\prime m}(\cos \theta) \sin \theta \Bigr)
  \hat e_\theta \\
  +\Bigl( -B_{lm} l P_{l-1}^{m}(\cos \theta) + C_{lm}\frac{(l+1)(l+2)}{2}
  P_{l+1}^{m}(\cos \theta) \Bigr) \hat e_r \Bigr]
\end{multline}

In the unbounded space, the leading terms have the order $l=0$ and
represent the 3 components of a Stokeslet. Their magnitude is given by the
coefficients $C_{0,-1}$, $C_{0,0}$ and $C_{0,1}$. In the next order, $l=1$,
representing terms that decay as $1/r^2$, we have 3 solutions for the $\chi$
component, 1 for $\Phi$ and 5 for $p$, 9 in total. In general, the number of
terms of order $l$ is
\begin{equation}
  N_l=3+6l\;.
\end{equation}

Note that all these solutions can be constructed from derivatives of 4
fundamental solutions. These include 3 Stokeslets ($l=0,m=0,\pm 1$) and a
source $l=1,m=0$. We denote these solutions as ${\cal G}_x$, ${\cal G}_y$,
${\cal G}_z$ and $\cal S$. The remaining solutions can be constructed from
derivatives like $\partial_x {\cal G}_x$, $\partial_y {\cal G}_x$ etc.

\subsection{Bounded by a no-slip plane}

As the cilium (or any other source of fluid pumping) is embedded in a plane,
we have to find the solutions that fulfill the boundary condition
\begin{equation}
  \vec{v}(r,\theta=\pi/2,\phi)=0\;.
\end{equation}
We first note that this condition has to be satisfied for all $r$ and
$\phi$ values, which means that we can collect terms with the same indices $l$
and $m$ and then form all independent linear combinations that fulfill the
boundary condition.

When expressing $\vec{v}_{lm}$ in the plane (at $\theta=\pi/2$) we have to distinguish
between even and odd values of $l+m$. If $l+m$ is even, so is the associated
Legendre polynomial $P_l^m(x)$, implying
$\left.\frac{d}{dx}P_l^m(x)\right|_{x=0}=0$.  For an odd $l+m$, $P_l^m(x)$ is
an odd function so that $P_l^m(0)=0$. Of course, we have to take into account
that the terms with an even $l+m$ actually contain $\Phi$ and $p$ terms with
an odd $l+m$, because they contain spherical harmonics of the order $l-1$ and
$l+1$, respectively. In total the condition that
$\vec{v}_{lm}(r,\pi/2,\phi)=0$ for an even $l+m$ implies
\begin{multline}
\label{eq:bcevenml}
im A_{lm} P_l^m(0) - B_{lm} P_{l-1}^{\prime m} (0)- \frac {1-l}{2} C_{lm}
P_{l+1}^{\prime m} (0) =0
\end{multline}
This condition is generally fulfilled by 2 independent solutions. For
example, we can choose the coefficients $A_{lm}$ and $C_{lm}$ freely, but then
have to determine $B_{lm}$ from (\ref{eq:bcevenml}). However, when $m=\pm l$,
there is a single solution, because there is no $\Phi$ ($B$) term. When
$l=m=0$ there is neither a $\Phi$ nor a $\chi$ term, so there is no solution
of this order.

For an odd $l+m$ the boundary condition leads to two equations
\begin{align}
\label{eq:bcoddml}
A_{lm} P_l^{\prime m}(0) + im B_{lm} P_{l-1}^{m}(0)+ im \frac {1-l}{2} C_{lm}
P_{l+1}^{m}&=0\nonumber\\
-l B_{lm} P_{l-1}^{m} (0) +  \frac {l^2+3l +2}{2} C_{lm}   P_{l+1}^{m}
  (0)&=0
\end{align}
These equations generally have a single solution for each pair $l$, $m$.

\begin{table}
\caption{Number of linearly independent solutions for each pair of numbers $l$
and $m$.}
\label{tab:1}       
\begin{tabular}{rllllll}
\hline\noalign{\smallskip}
$m$   & 0 & $\pm 1$ & $\pm 2$ & $\pm 3$ & $\pm 4$ \ldots & Total\\
\noalign{\smallskip}\hline\noalign{\smallskip}
$l=0$ & 0 &        &         &         &                & 0 \\
1     & 1 & 1      &         &         &                & 3 \\ 
2     & 2 & 1      & 1       &         &                & 6 \\
3     & 1 & 2      & 1       &  1      &                & 9 \\
4     & 2 & 1      & 2       &  1      &  1             & 12\\
$\vdots$&&&&&&$\vdots$\\
$l$ &&&&&& $3l$ \\
\noalign{\smallskip}\hline
\end{tabular}
\end{table}
It is now time to count the total number of independent solutions for each
$l$. An overview is given in Table~\ref{tab:1}. For any $l\ge 1$, we have
\begin{equation}
  \label{eq:numbersolutions}
  N_l=3 l 
\end{equation}
solutions, down from $3+6l$ in the unbounded space. 

In the following we will have a closer look at the terms that decay
quadratically with the distance ($l=1$) and those that decay with the third
power ($l=2$).

\subsection{\boldmath $l=1$: Modes of the order $r^{-2}$}

In this order we have $N_1=3$ modes: one of them ($m=0$) represents a fluid
source and two ($m=\pm 1$) the flow of two horizontal Stokeslets in the
proximity of the boundary. 

The source is described by the second and the fourth term in Eq.~(\ref{eq:v})
alone. The solution for the pressure is $p_{10} \propto \frac 1 {r^3} \frac {3
  \cos^2\theta -1}{2}$, that for the second term $\Phi_{00} \propto 1/r$.
The resulting velocity is
\begin{equation}
  \label{eq:source}
  S(\vec{x})=\frac 1 {r^2} \cos^2 \theta
  \hat e_r = \frac {z^2}{r^5}\left( \begin{array}{c}x\\y\\z\end{array} \right)\;.
\end{equation}
To distinguish it from a source in bulk we call this term \textit{``surface
  sourcelet''}.  Note that because of volume conservation beating cilia
naturally do not act as a source. But the source is one of the fundamental
singularities and we will later show that its derivatives can be used to
express higher order terms that are present. Also, this formalism could be
used in a more general context -- the source could describe fluid injection
through a pore in the surface, or, temporarily, the flow around an expanding
bubble.

The solutions with $m=\pm 1$ represent two horizontal Stokeslets.  In this
order there is no $\Phi$ term and the $\chi$ term has to vanish ($A_{lm}=0$)
in order to fulfill the boundary condition~(\ref{eq:bcevenml}). The velocities
are determined by the last term in Eq.~(\ref{eq:v}) and stem from pressure
terms $p_{2,\pm 1}\propto \frac 1 {r^3} e^{\pm i m \phi} \cos \theta \sin
\theta$. We can construct the first from $\vec{v}_{1,1}+\vec{v}_{1,-1}$
\begin{equation}
  D_x=\frac 1 {r^2} \cos \theta \sin \theta \cos \phi \hat e_r =  \frac
  {xz}{r^5}\left( \begin{array}{c}x\\y\\z\end{array} \right)
\end{equation}
and the second from $\vec{v}_{1,1}-\vec{v}_{1,-1}$
\begin{equation}
  D_y=\frac 1 {r^2} \cos \theta \sin \theta \sin \phi \hat e_r =  \frac
  {yz}{r^5}\left( \begin{array}{c}x\\y\\z\end{array} \right)\;.
\end{equation}
We will call these terms \textit{``surface Stokeslets''}. With 3 linearly
independent solutions we know that we have a complete set for the order
$v\propto r^{-2}$.

\begin{figure}[t!]
\begin{center}
$S$     \includegraphics[width=8cm]{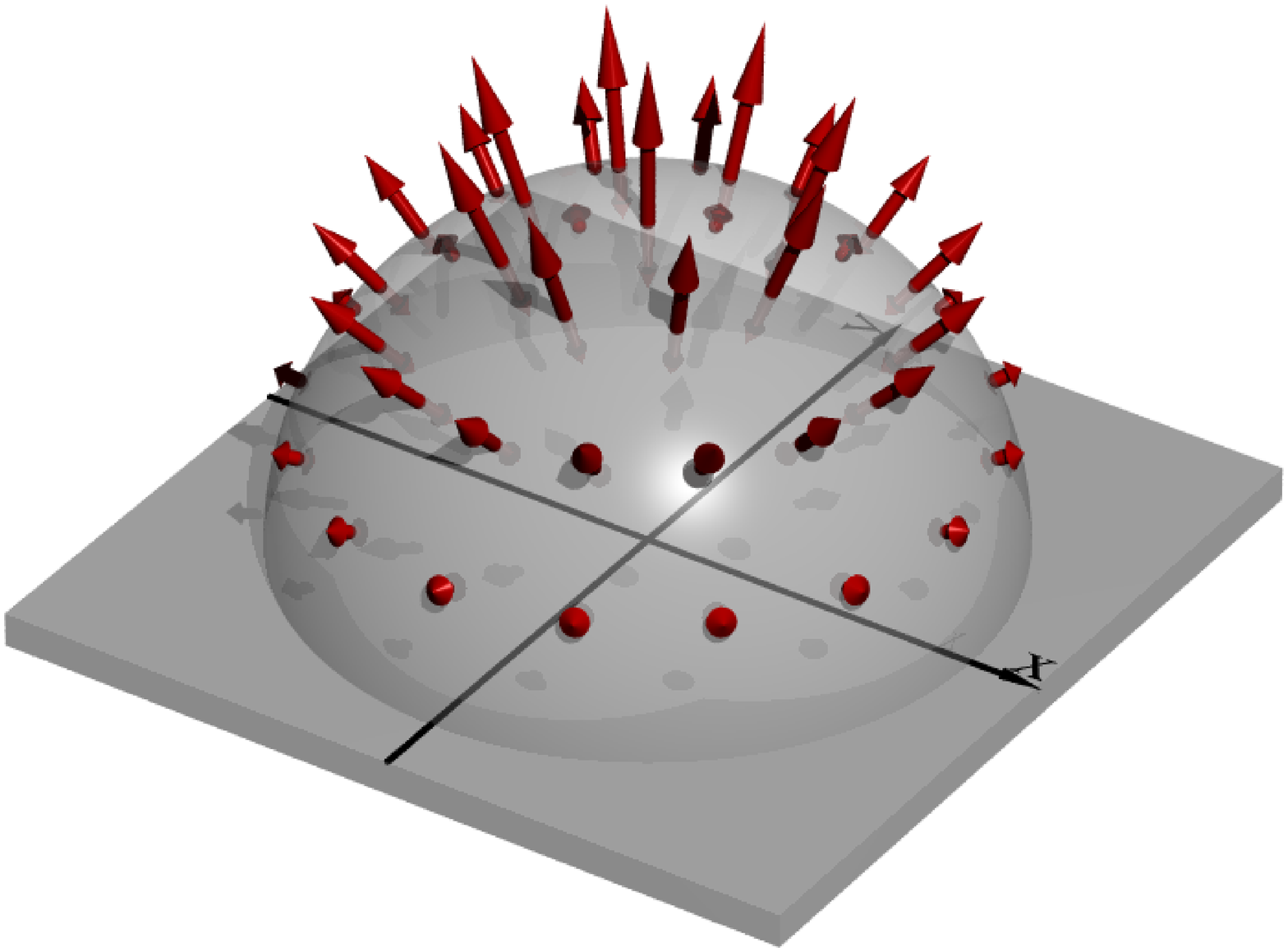}\\
$D_x$   \includegraphics[width=8cm]{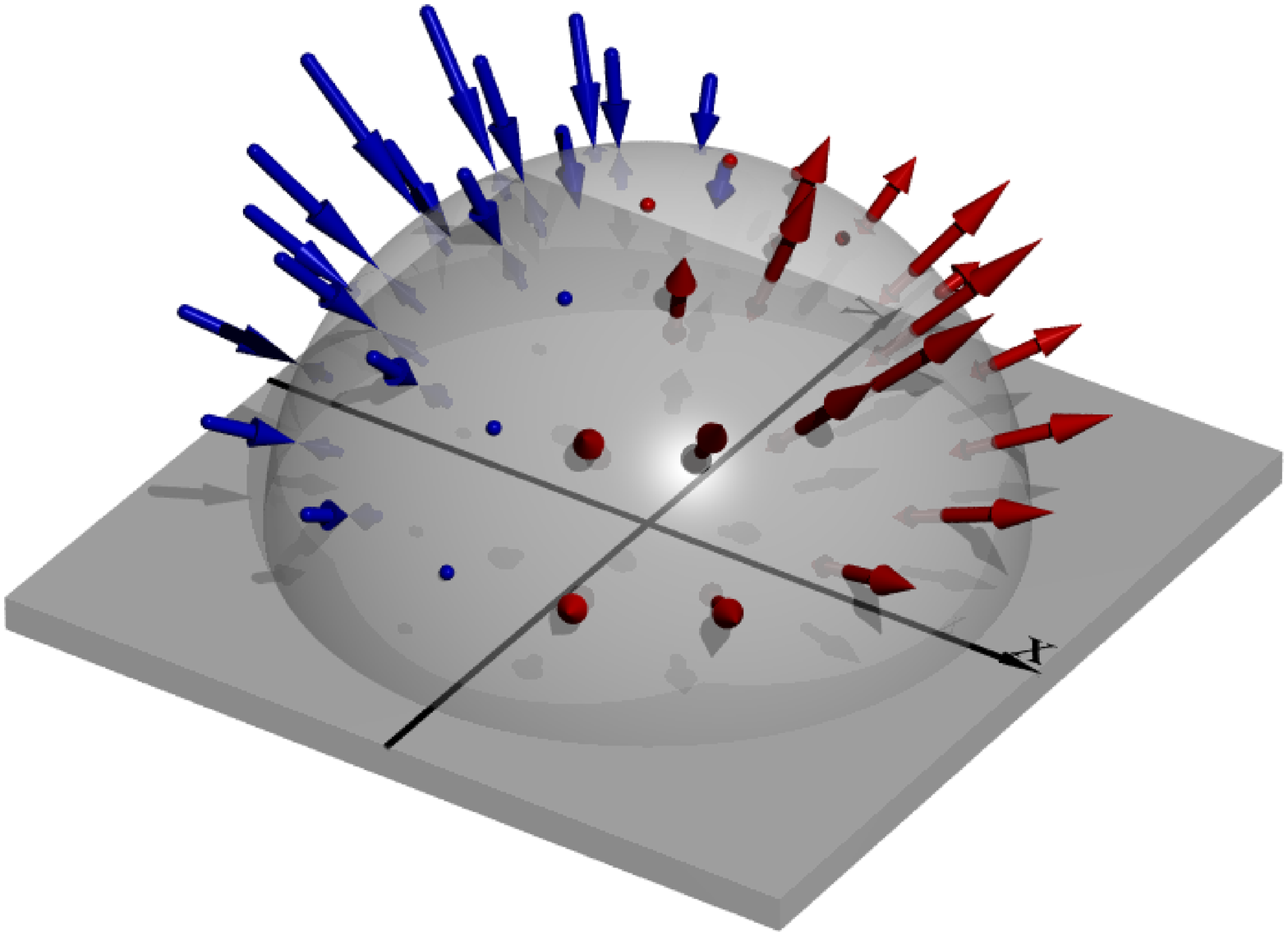}\\
$D_y$   \includegraphics[width=8cm]{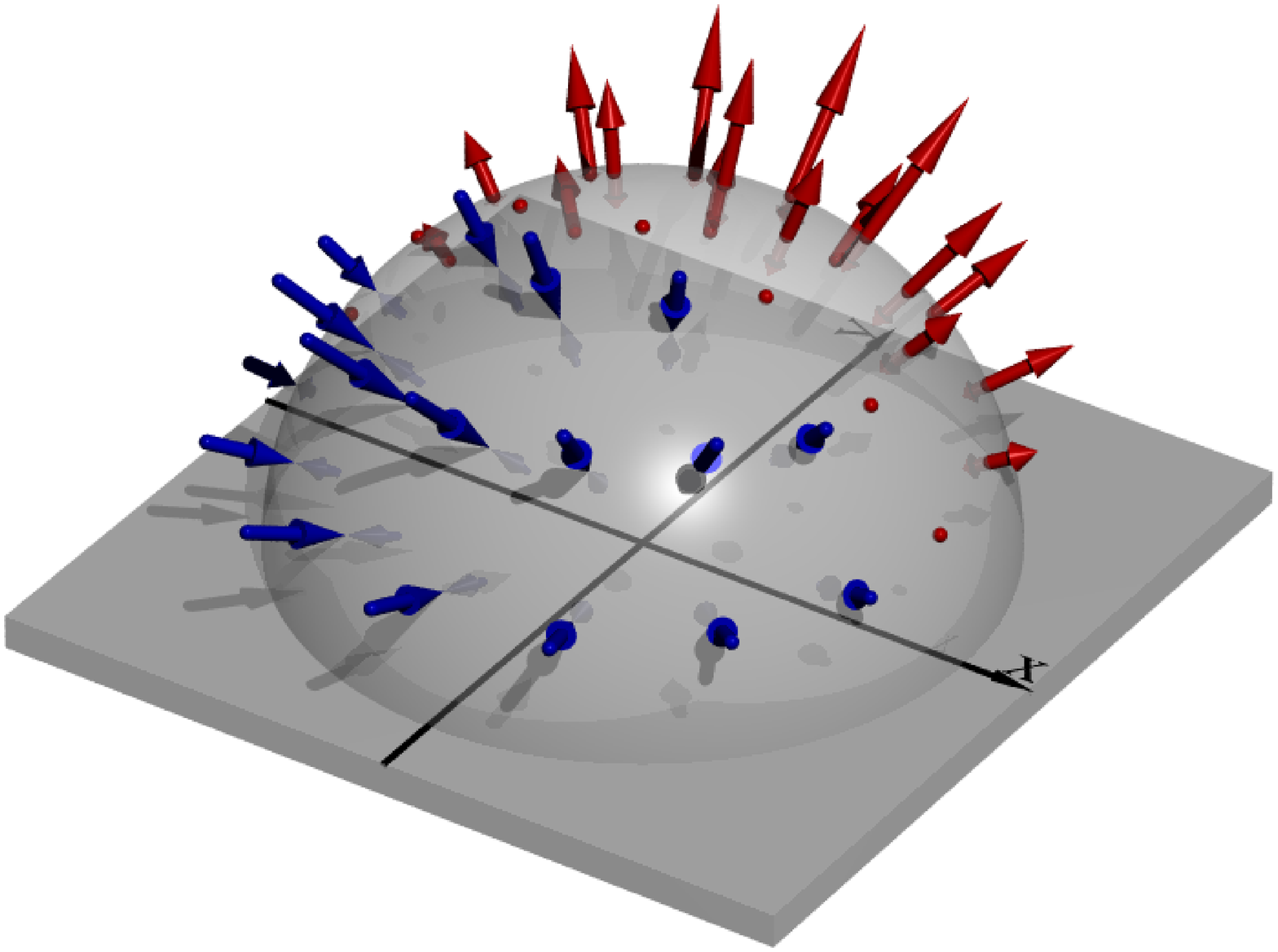}
 \end{center}
 \caption{Fundamental modes of the order $r^{-2}$. For each mode the angular
  dependence of the velocity field is shown along a semi-transparent
  hemisphere. The colour of the arrows denotes inbound and outbound flow. The
  three depicted modes are a ``surface sourcelet'' $S$ and two ``surface
  Stokeslets'' $D_x$ and $D_y$.}
\label{fig:1}
\end{figure}

The surface sourcelet $S$ and the two surface Stokeslets $D_x$ and $D_y$ are
shown in Figure \ref{fig:1}. 
Note that these 3 modes represent a fundamental set. All higher order terms
can be constructed by deriving them by $x$ and $y$. We can see this from the
fact that we have counted $3l$ solutions of order $l$. To get from the
fundamental solutions to those of order $l$ we have to derive them $l-1$ times
by the coordinates, which gives us $l$ linearly independent derivatives
(unlike in unbounded fluid they all are independent). So we have shown that
in total $3l$ derivatives of the fundamental solutions represent a complete set
of solutions to that order. We will nevertheless derive the solutions from the
spherical harmonics directly in order to collect modes according to their
angular symmetry - this system is very convenient to discuss the ciliary
flows. 

\subsection{\boldmath $l=2$: Modes of the order $r^{-3}$}

\begin{figure*}
\begin{center}
 $Q_1$   \includegraphics[width=8cm]{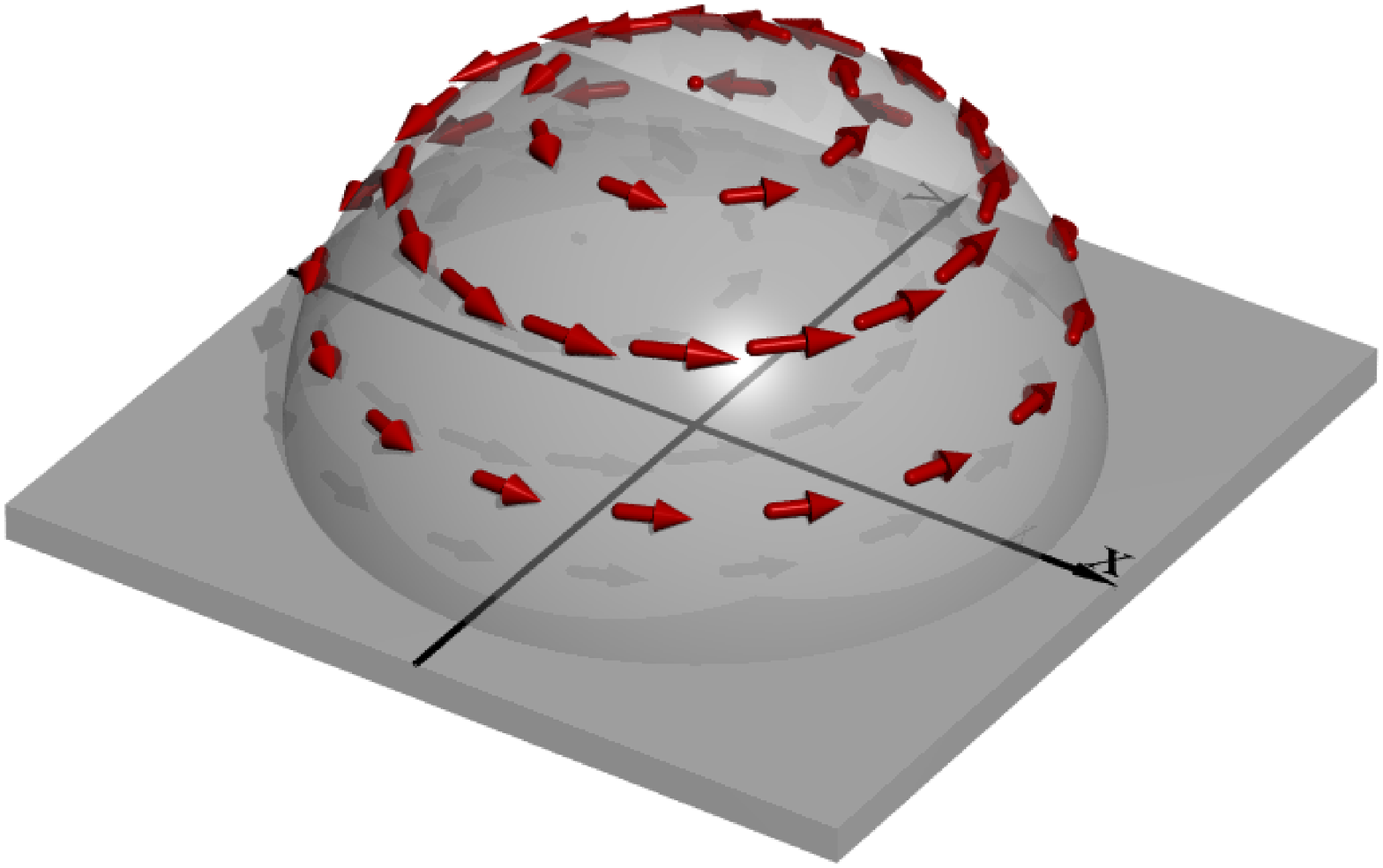} 
 $Q_2$   \includegraphics[width=8cm]{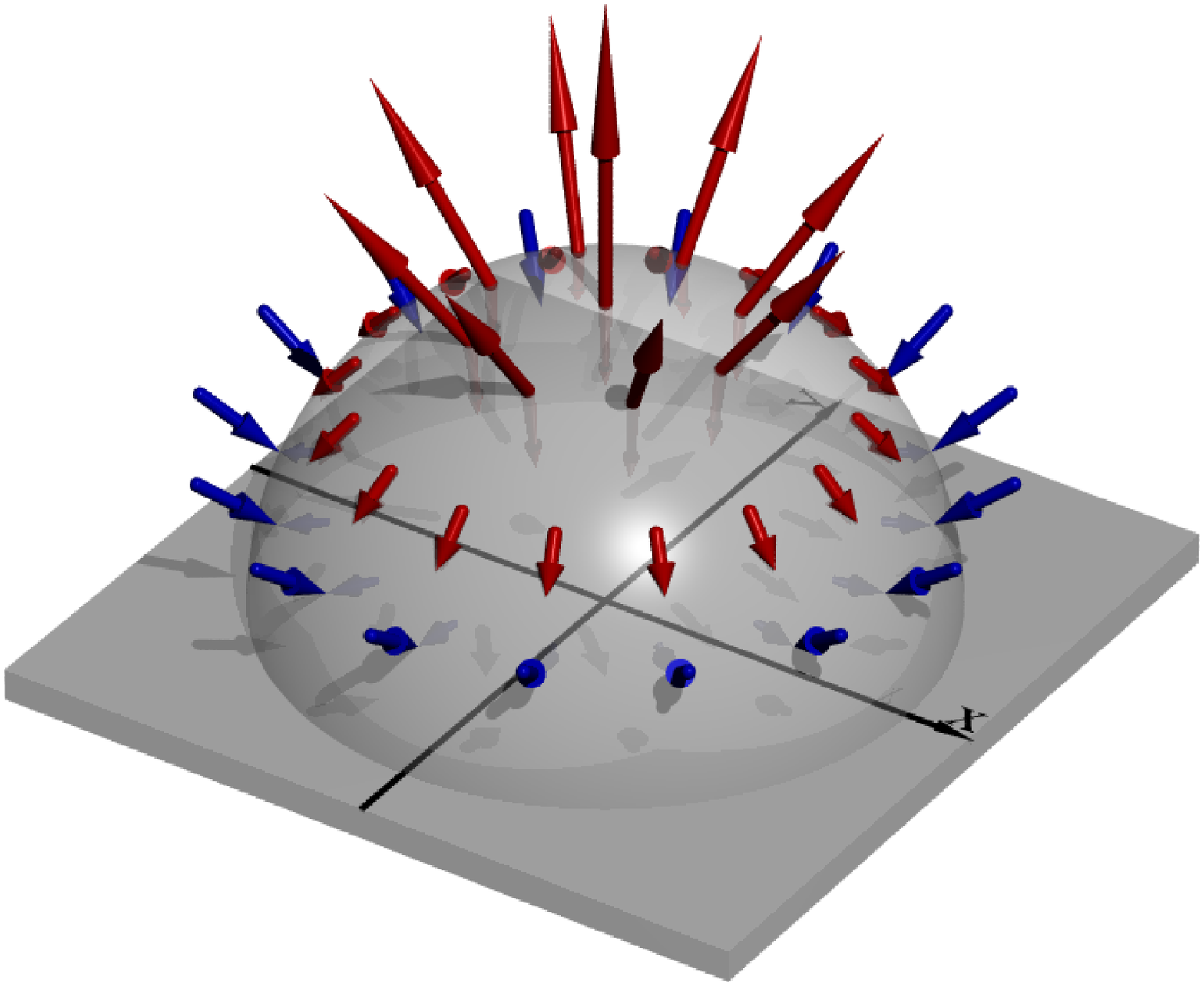}\\
 $Q_3$   \includegraphics[width=8cm]{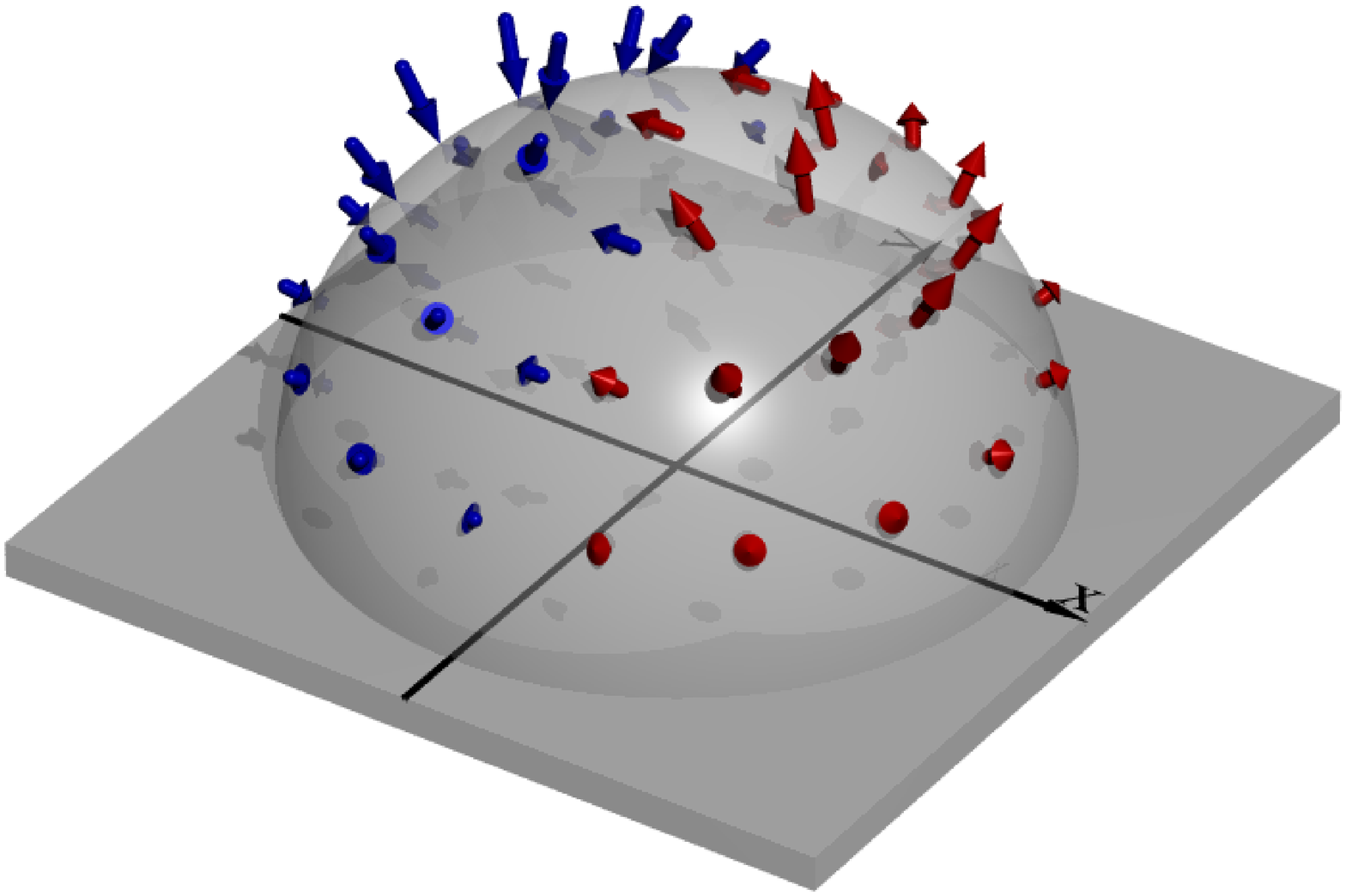}
 $Q_4$   \includegraphics[width=8cm]{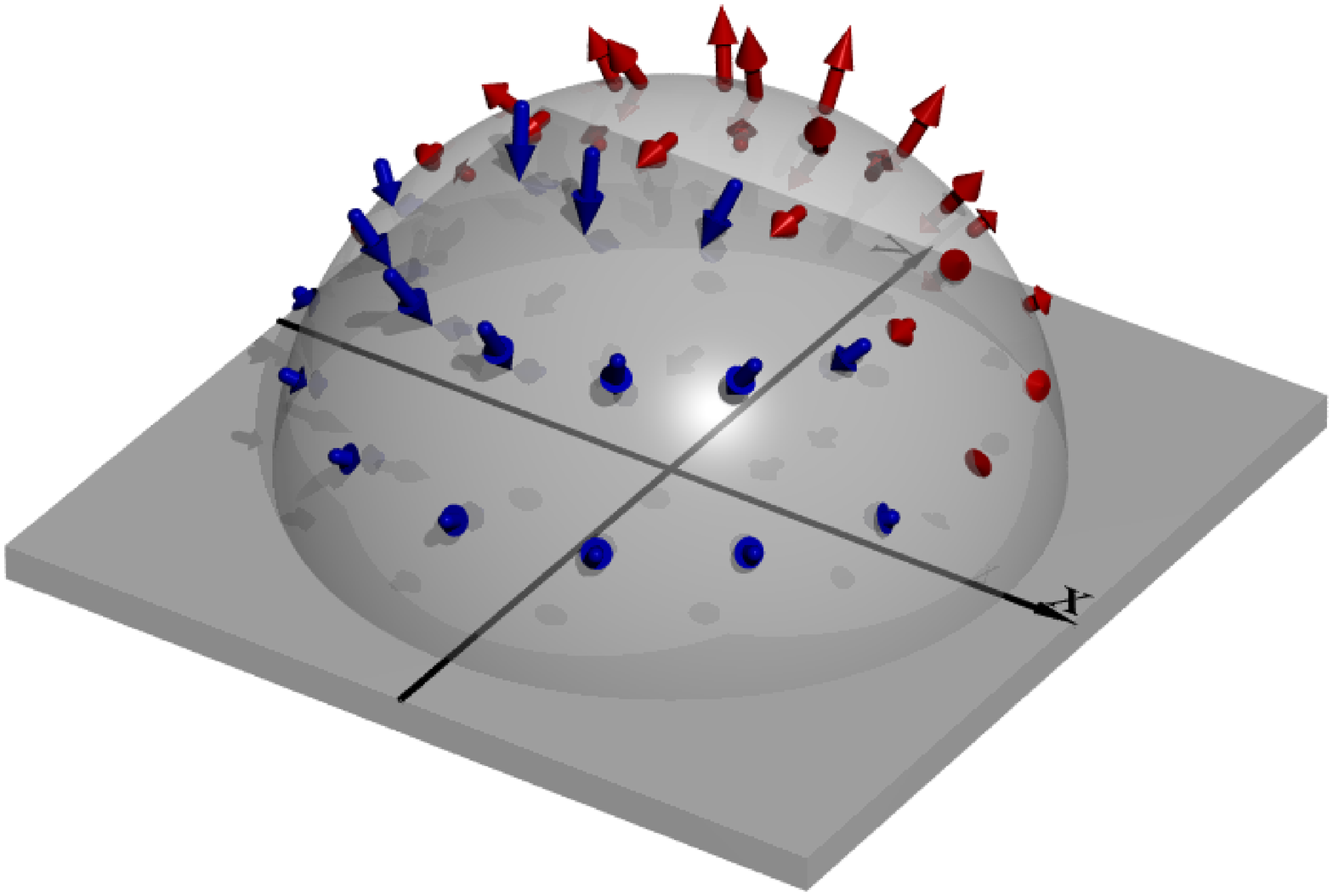}\\
 $Q_5$   \includegraphics[width=8cm]{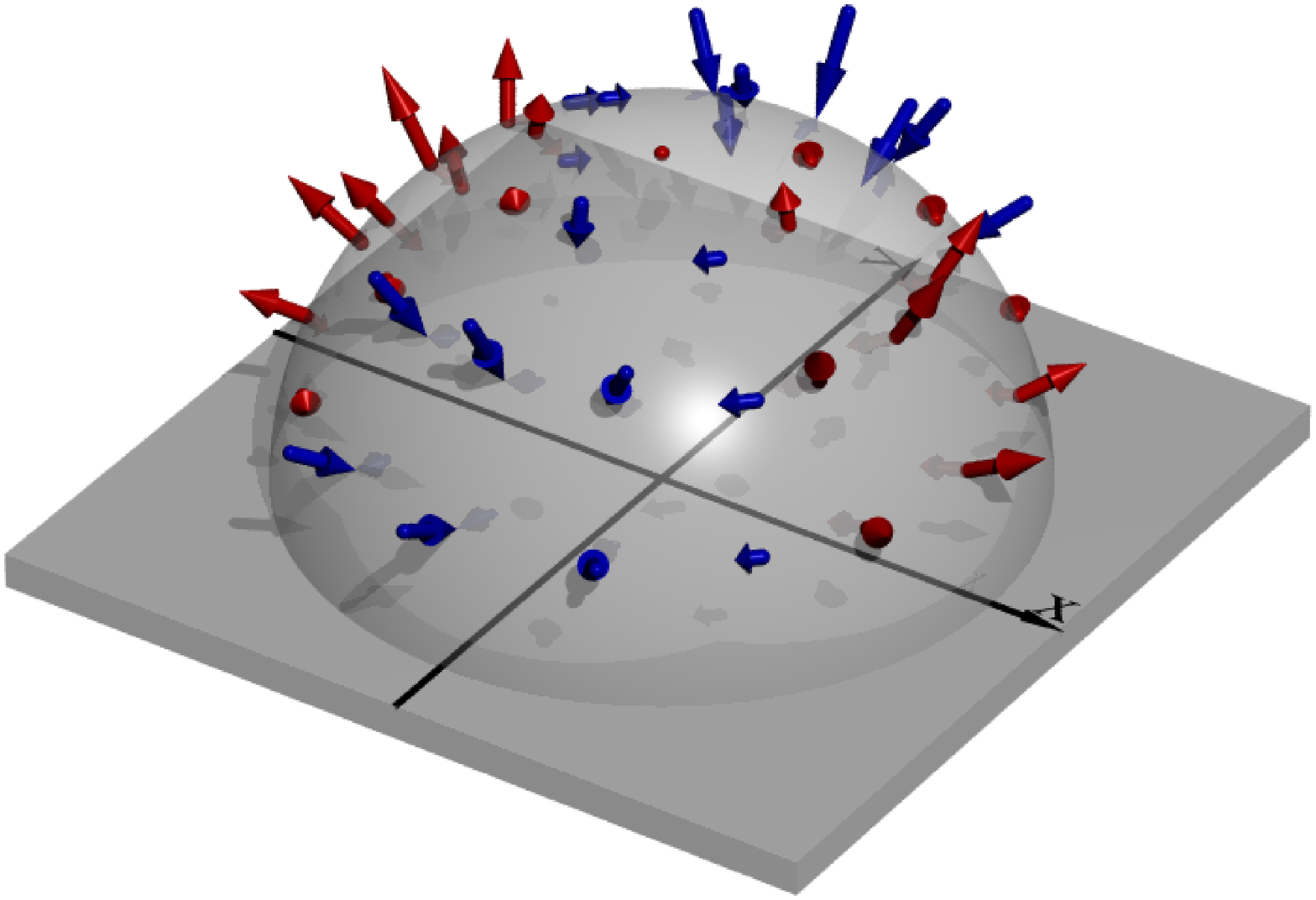}
 $Q_6$   \includegraphics[width=8cm]{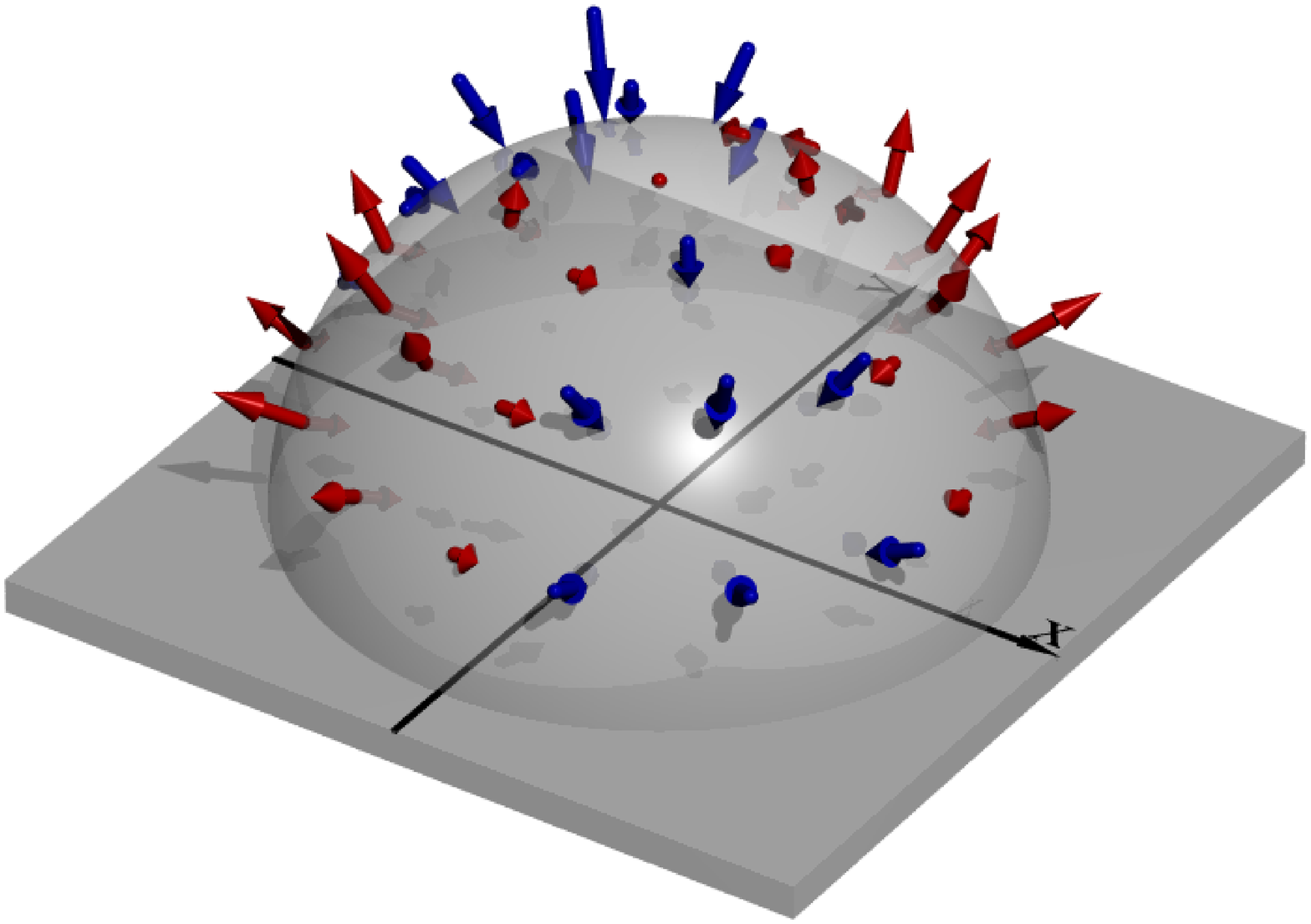}
 \end{center}
 \caption{Six modes of the order $r^{-3}$: The surface rotlet $Q_1$, the
   vertical Stokeslet $Q_2$, the surface source doublets $Q_3$ and $Q_4$ and
   surface stresslets $Q_5$ and $Q_6$.}
\label{fig:2}
\end{figure*}

In this order we have a total of 6 independent terms, two with $m=0$, two with
$m=\pm 1$ and two with $m=\pm 2$. 

The first solution can be obtained from $\chi_{2,0}\propto \frac 1
{r^3}\frac{3 \cos^2 \theta -1}{2}$:
\begin{equation}
  Q_1=\frac 1 {r^3} \sin \theta \cos \theta \hat e_\phi =  \frac
  {z}{r^5}\left( \begin{array}{c}-y\\x\\0\end{array} \right)\;.
\end{equation}
This is a rotlet around the $z$ axis. 

The second solution is obtained from $\Phi_{1,0}$ and $p_{3,0}$. It reads
\begin{align}
  Q_2&=\frac 1 {r^3} \left( (4\cos^3\theta -2 \cos \theta) \hat{e}_r + \cos^2
    \theta \sin \theta \hat{e}_\theta \right)\nonumber \\&=  \frac
  {z}{r^7}\left( \begin{array}{c}(5z^2-2r^2)x\\ (5z^2-2r^2) y \\ (5z^2-3r^2) z \end{array} \right)\;
\end{align}
and corresponds to the flow of a vertical Stokeslet near the boundary plane
\cite{Blake.Chwang1974}. The fluid is moved outwards along the $z$ axis and
inwards along the $x-y$ plane. Note that this is the only axisymmetric term --
it is therefore clear that this is the leading term describing acoustic
streaming caused by ultrasonic oscillations of a small bubble on a planar
surface \cite{Marmottant.Hilgenfeldt2006}.

The modes with $m\pm1$ that fulfill the boundary condition (\ref{eq:bcoddml})
are
\begin{align}
 Q_3&=\Re \frac{1}{r^{3}} \left( 4 \cos^2 \theta \sin \theta \hat{e}_r - \cos^3
  \theta  \hat{e}_\theta - i \cos^2 \theta \hat{e}_\phi\right) e^{i\phi} \nonumber\\&=
\frac{z^2}{r^{7}} \left( \begin{array}{c}
5  x^2- r^2  \\
5  xy\\
5  xz
\end{array}\right)\\
Q_4&=\Im \frac{1}{r^{3}} \left( 4 \cos^2 \theta \sin \theta \hat{e}_r - \cos^3
  \theta  \hat{e}_\theta - i \cos^2 \theta \hat{e}_\phi\right)  e^{i\phi} \nonumber\\&=
\frac{z^2}{r^{7}} \left( \begin{array}{c}
5  xy  \\
5  y^2-r^2\\
5  yz
\end{array}\right)
\end{align}
They represent the fields of a horizontal rotlet (around the $y$ and $x$ axes)
in the presence of a boundary. As we will see later, they can also be viewed
as source-sink pairs (source-doublets) on the surface. 

Finally, the modes with $m=\pm 2$, which have to fulfill the boundary
condition (\ref{eq:bcevenml}) are
\begin{align}
  Q_5&=\Re \frac 1 {r^3} \sin \theta \cos \theta \left( 4 \sin \theta\hat{e}_r
    - \cos\theta \hat{e}_\theta - i \hat{e}_\phi
  \right) e^{ 2 i \phi}\nonumber \\
  &=\frac{z}{r^7} \left( \begin{array}{c}
      (5 (x^2-y^2)-r^2) x\\
      (5 (x^2-y^2)+r^2) y\\
      5(x^2-y^2) z \\
  \end{array}
\right)\\
Q_6&=\Im \frac 1 {r^3} \sin \theta \cos \theta \left( 4 \sin\theta \hat{e}_r -
  \cos \theta \hat{e}_\theta - i \hat{e}_\phi
\right) e^{ 2 i \phi}\nonumber \\
&=\frac{z}{r^7} \left( \begin{array}{c}
    (10x^2-r^2) y\\
    (10y^2-r^2) x\\
    10xy  z \\
  \end{array}
  \right)
\end{align}
These modes represent surface stresslets. $Q_5$ is generated if a pair of
forces pulls the fluid apart along the $x$ axis and another pair together
along the $y$ axis. $Q_6$ is similar, but rotated by $45^\circ$ about the
$z$ axis.

\begin{figure}
\begin{center}
   \includegraphics{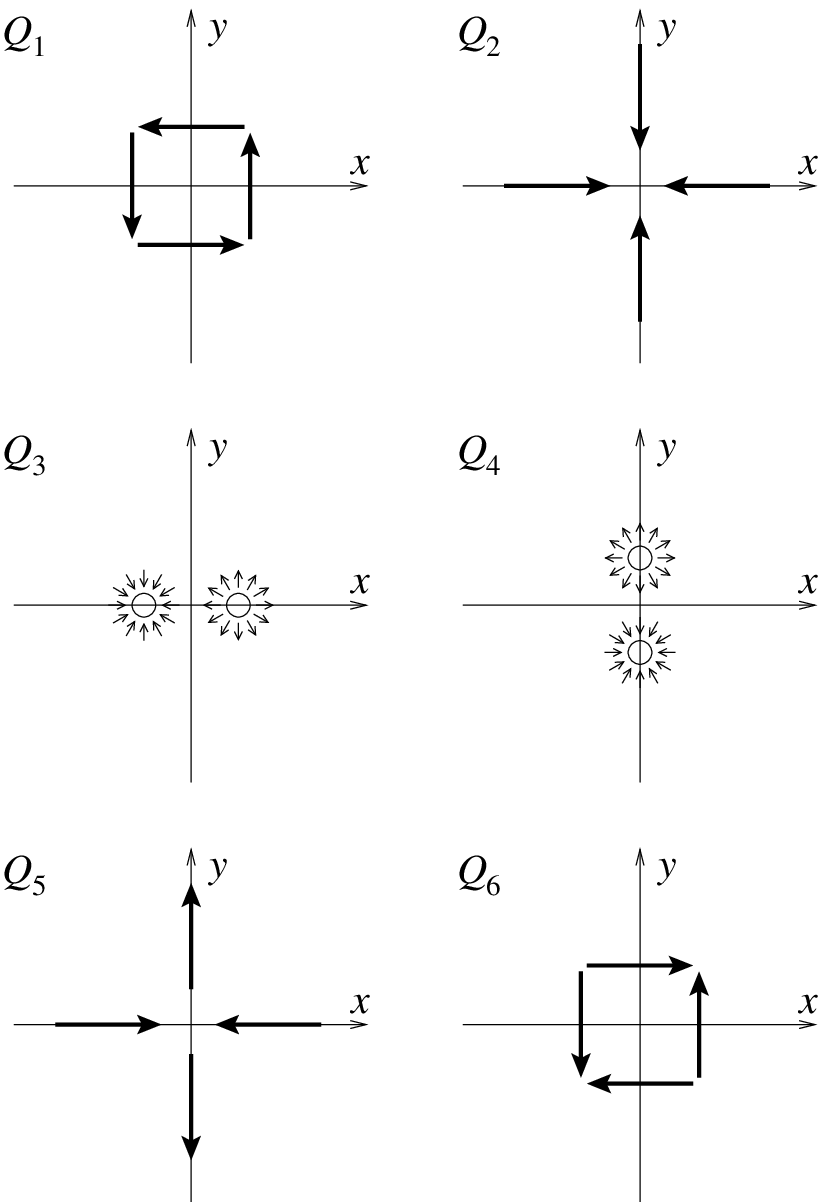} 
 \end{center}
 \caption{Six modes of the order $r^{-3}$ represented with fundamental
   $r^{-2}$ solutions.}
\label{fig:3}
\end{figure}

Like in unbounded space \cite{Pozrikidis1992} all these modes can be
represented by derivatives of the $\sim r^{-2}$ modes in the following way:
\begin{align}
\label{eq:d1deriv}
  Q_1&=\frac{\partial D_x}{\partial y} - \frac{\partial D_y}{\partial x}\\
  Q_2&=\frac{\partial D_x}{\partial x} + \frac{\partial D_y}{\partial y}\\
  Q_3&=-\frac{\partial S}{\partial x}\\
  Q_4&=-\frac{\partial S}{\partial y}\\
  Q_5&=-\frac{\partial D_x}{\partial x} + \frac{\partial D_y}{\partial y}\\
  Q_6&=-\frac{\partial D_x}{\partial y} - \frac{\partial D_y}{\partial x}
\label{eq:d6deriv}
\end{align}
These relations indicate that the modes $Q_1$, $Q_2$, $Q_5$ and $Q_6$ can be
seen each as the flow of four point forces at an infinitesimal distance, as
well as infinitesimally close to the surface. $Q_3$ and $Q_4$ represent a
dipole consisting of a source and a sink. This representation is illustrated
in Fig.~\ref{fig:3}.

\subsection{Higher order modes}

In each higher order $l$, we have $N_l=3l$ modes that fall off with the power
$r^{-(l+1)}$. All of them can be constructed from the derivatives
\begin{align}
  \frac {\partial^{l-1}S}{\partial x^j \partial y^{l-1-j}}\;, \qquad
  \frac {\partial^{l-1}D_x}{\partial x^j \partial y^{l-1-j}}\;,\; \text{and} \qquad
  \frac {\partial^{l-1}D_y}{\partial x^j \partial y^{l-1-j}}\;.
\end{align}
Unlike in unbounded fluid all these derivatives are linearly independent. 

\subsection{Normal form of the flow profile}
\label{sec:normalform}

Ciliary flows are characterized by the fact that they do not contain a source,
but they do generate directional flow, so at least one of the terms $D_x$ and
$D_y$ is present. By rotating the coordinate system, we can always bring the
time-averaged flow to the form $\bar {\vec{v}}(\vec{x})={\cal A} D_x +{\cal
  O}(r^{-3})$.

To next order, the flow now has the shape 
\begin{equation}
  \bar {\vec{v}}(\vec{x})={\cal A} D_x(\vec{x}) +\sum_{i=1}^6 {{\cal B}_i}
  D_i(\vec{x}) +{\cal
    O}(r^{-4})\;.
\end{equation}
If we transform the coordinate system to $\vec{x'}=\vec{x}-d_x \hat e_x - d_y
\hat e_y$, we have 
\begin{multline}
  \bar {\vec{v}}'(\vec{x}')=\bar{\vec{v}}(\vec{x}'-d_x \hat e_x - d_y
    \hat e_y) \\= {\cal A} \left(D_x(\vec{x'}) - d_x \frac {\partial D_x}{\partial
        x}(\vec{x'})- d_y \frac {\partial D_x}{\partial
        y}(\vec{x'}) \right) \\+ \sum_{i=1}^6 {{\cal B}_i}
  D_i(\vec{x}') +{\cal
    O}(r^{-4})\;.
\end{multline}
As we know from Eqns.~(\ref{eq:d1deriv}-\ref{eq:d6deriv}) the derivatives are
\begin{align}
  \frac {\partial D_x}{\partial x} &= \frac{Q_2-Q_5}{2} &   \frac {\partial
    D_x}{\partial y} &= \frac{Q_1-Q_6}{2}
\end{align}
and by choosing $d_x=-\frac {2 {\cal B}_5}{\cal A}$ and $d_y=-\frac {2 {\cal
  B}_6}{\cal A}$ we can eliminate the modes $Q_5$ and $Q_6$. Of course, this
transformation is only possible if ${\cal A}\ne 0$.  Thus we have brought the
stationary flow to the normal form
\begin{equation}
  \bar {\vec{v}}'(\vec{x}')= {\cal A} D_x(\vec{x'}) + \sum_{i=1}^4 {{\cal B}'_i} D_i(\vec{x}') +{\cal
    O}(r^{-4})\;.
\end{equation}

To summarize, with the proper choice of the coordinate system the static flow
generated by a cilium can be described up to the order $r^{-3}$ with only 5
components: the surface Stokeslet $D_x$, the surface rotlet $Q_1$, the
vertical surface Stokeslet $Q_2$, and the surface source-doublets $Q_3$ and
$Q_4$. Because we can only transform the coordinate system once, for the
static flow, we cannot reduce the oscillatory flows in the same way. 

\begin{figure}
  \begin{center}
    \includegraphics[height=6cm]{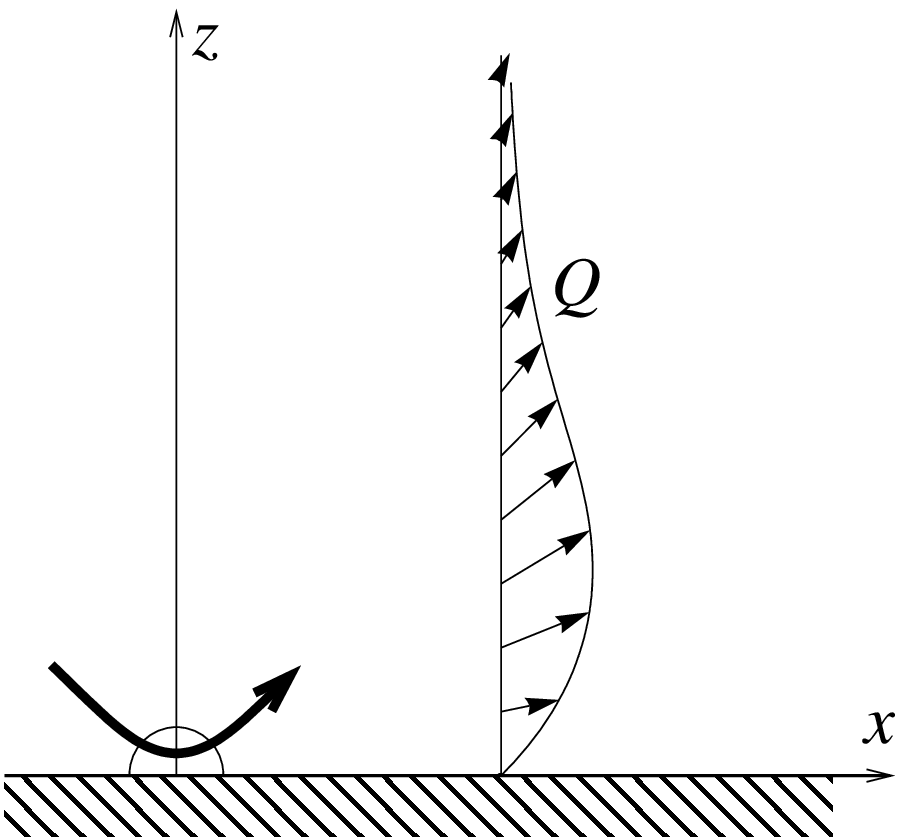}\\[1cm]
    \includegraphics[height=6cm]{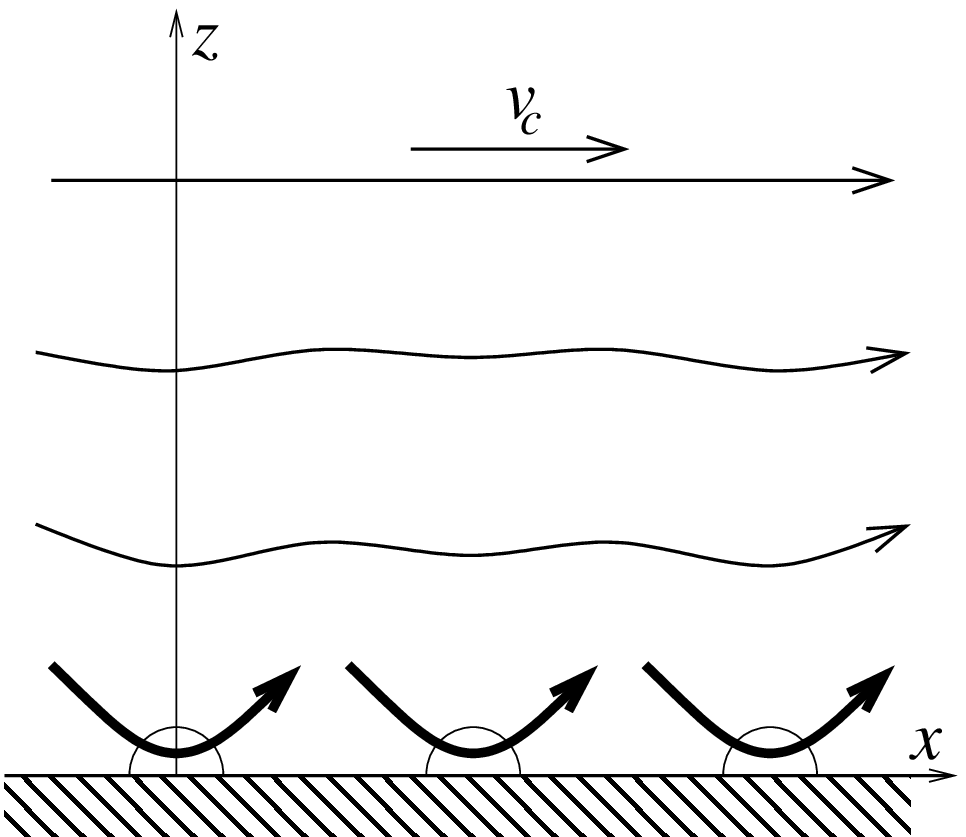}
  \end{center}
  \caption{\label{fig:crossection}a) The volume flow rate $Q$ is defined as
    the fluid flux through a half-plane perpendicular to the surface and to
    the direction of pumping. b) Above a densely ciliated surface the flow
    becomes homogeneous with the average velocity $v_c$.}
\end{figure}

\subsection{Volume flow rate}

The performance of a cilium is characterized by the volume flow rate $Q$,
defined as the average flux through a half-plane perpendicular to the
direction of pumping \cite{Smith.Gaffney2008},
Fig.~\ref{fig:crossection}a. For example, for the mode corresponding to the
pumping in $x$-direction with amplitude $\cal A$, $\vec v(x,y,z)={\cal A}
D_x(x,y,z)$ has the volume flow rate
\begin{equation}
  \label{eq:volflowcalc}
  Q=\int_{-\infty}^\infty dy \int_0^\infty dz\, v_x (x,y,z)=\frac 2 3 {\cal A}\;.
\end{equation}
Due to volume conservation the flow rate is, of course, independent of the
$x$-position of the cross-section chosen. Similarly, the volume flow rate of
the mode $D_y$ would be defined as the flux through a half-plane with a
constant $y$.

For the source, the flow rate is the actual influx. For a flow with $\vec
v(x,y,z)={\cal A} S(x,y,z)$, the flow rate is
\begin{equation}
  Q=\frac {2\pi}{3} {\cal A}\;.
\end{equation}

The volume flow rate for higher order modes ($l\ge 2$) is always zero - the
easiest way to see this is from the fact that they can be represented as
spatial derivatives of the fundamental modes.

\subsection{Velocity above ciliated layer}

Instead of a single cilium, we often deal with a densely ciliated surface
(Fig.~\ref{fig:crossection}b). Let $\rho$ denote the surface density of those
cilia. Then the velocity above an infinite array is
\begin{equation}
  \label{eq:vinfty}
  v_c=\int_{-\infty}^\infty dX \int_{-\infty}^\infty dY \rho
 {\cal A}  D_x(x-X,y-Y,z)= \frac {2\pi}{3} \rho {\cal A} = \pi \rho Q\;,
\end{equation}
which is independent of $z$. The second relationship expresses the velocity
above a ciliated layer with the volume generated by each cilium. In this
regime, one can simplify the description of cilia by replacing them with a
surface slip term with velocity $v_c$ \cite{Julicher.Prost2009}.

\section{Ciliary models}

The ciliary beating pattern is quite complex, but there are several simplified
models that have the correct symmetry properties and capture the far-field
flow patterns. The simplest of all is a model that replaces the cilium with a
single small particle (radius $a$), moving periodically on a closed path, and
was used in several models for ciliary synchronization
\cite{vilfan2006a,Uchida.Golestanian2010,Golestanian.Uchida2011,Uchida.Golestanian2011,Kotar.Cicuta2010}.

\subsection{Cilium as a small sphere}

We start our discussion by expanding the flow field of a small particle
subject to force $\vec{F}$ in terms of $D$- and $Q$-modes. An exact solution
for the flow has been derived by Blake \cite{Blake.1971} and is now sometimes
referred to as \textit{Blake's tensor}:
\begin{multline}
  \label{eq:blake}
  \vec{v}_\alpha(\vec{x}) = \sum_\beta  \bigl[ G_{\alpha \beta}^S (\vec{x}-\vec{R}) -
  G_{\alpha \beta}^S (\vec{x}-\vec{\bar R}) + 2 Z^2
    G_{\alpha \beta}^D (\vec{x}-\vec{\bar R}) \\-2Z  G_{\alpha \beta}^{SD}
    (\vec{x}-\vec{\bar R})
  \bigr] F^\beta
\end{multline}
where $\vec{R}=(X,Y,Z)$ is the position of the particle, $Z$ is its
height above the plane, $\vec{\bar R}=(X,Y,-Z)$ is the position of its
mirror image and
\begin{align}
  G_{\alpha \beta}^S(\vec{x})&=\frac 1 {8 \pi \eta}\left( \frac{\delta_{\alpha
  \beta}}{\left| \vec{x} \right|} +  \frac{{x}_{\alpha}{x}_{
  \beta}}{\left| \vec{x} \right|^3} \right)\;,\\
 G_{\alpha \beta}^D (\vec{x})&= \frac{1}{8 \pi \eta} (1-2\delta _{\beta z})
  \frac{\partial}{\partial
    x_\beta} \left( \frac{x_\alpha}{\left| \vec{x} \right|^3} \right)\\
\intertext{and}
  G_{\alpha \beta}^{SD} (\vec{x})&=(1-2\delta _{\beta z})
  \frac{\partial}{\partial x_\beta} G_{\alpha z}^S(\vec{x})
\end{align}
are the fields of a Stokeslet, source doublet and a Stokeslet-doublet,
respectively. A Taylor expansion of Blake's expression for small particle
located at $(0,0,Z)$ up to the order $Z^2$ reads
\begin{multline}
  \vec{v}(\vec{x})=\frac 1 {8\pi \eta} \bigl( 12 Z \bigl( F_x D_x(\vec{x})+
      F_y D_y(\vec{x})\bigr)\\ + 6 Z^2\bigl(  - F_x Q_3(\vec{x}) 
 - F_y Q_4(\vec{x}) +  F_z Q_2(\vec{x}) \bigr)\bigr)\;.
\end{multline}
As noted by Blake and Chwang \cite{Blake.Chwang1974} the horizontal forces
$F_x$ and $F_y$ generate terms of the order $r^{-2}$, while the vertical force
$F_z$ only generates terms $\sim r^{-3}$.  If the particle is not located
exactly on the $z$ axis, but in any point $(X,Y,Z)$, we can use the
derivatives (\ref{eq:d1deriv}--\ref{eq:d6deriv}) to obtain an expression that
is again exact up to the order $r^{-3}$:
\begin{align}
\label{eq:velocity-gen}
\vec{v}(\vec{x})&=\frac 1 {8\pi \eta} \bigl( 12 Z \bigl( F_x D_x(\vec{x})+ F_y
D_y(\vec{x})\bigr)\nonumber\\ &+ 6 Z^2\bigl( - F_x Q_3(\vec{x})
- F_y Q_4(\vec{x}) +  F_z Q_2(\vec{x}) \bigr)\nonumber\\
&+6 ZX F_x (Q_5(\vec{x})- Q_2(\vec{x})) \nonumber\\&+6 ZY F_x (Q_6(\vec{x})-
Q_1(\vec{x}))\nonumber\\
&+6 ZX F_y (Q_6(\vec{x})+ Q_1(\vec{x}))\nonumber\\& + 6 ZY F_y  (-Q_5(\vec{x})-
Q_2(\vec{x}))
\bigr)\;.
\end{align}
This equation provides a basis for discussing all models that replace the
cilium with a point particle. 

\begin{figure*}
\begin{center}
  \begin{tabular}{ccc}
    a)\includegraphics[width=0.3\textwidth]{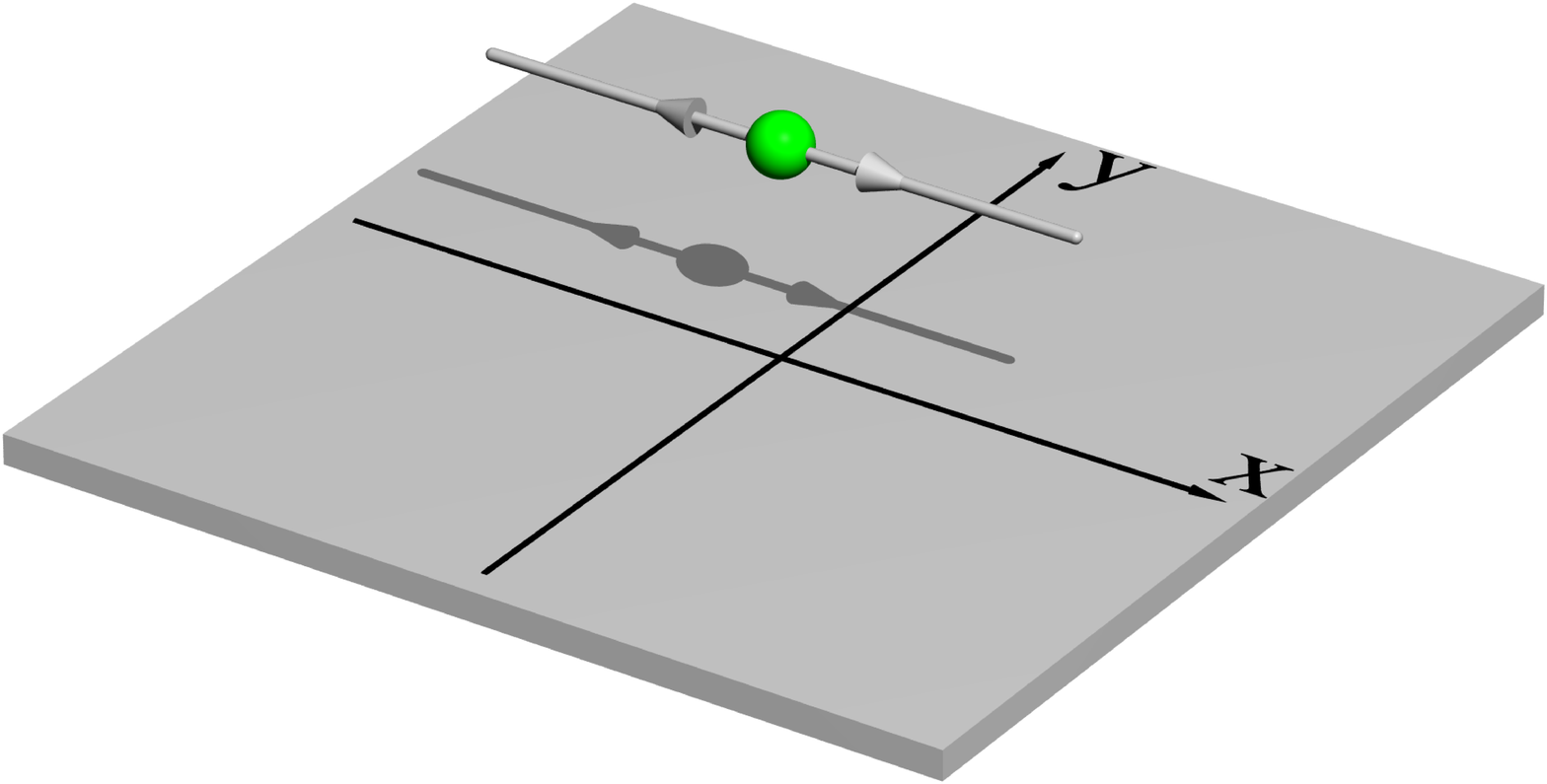}&
    b)\includegraphics[width=0.3\textwidth]{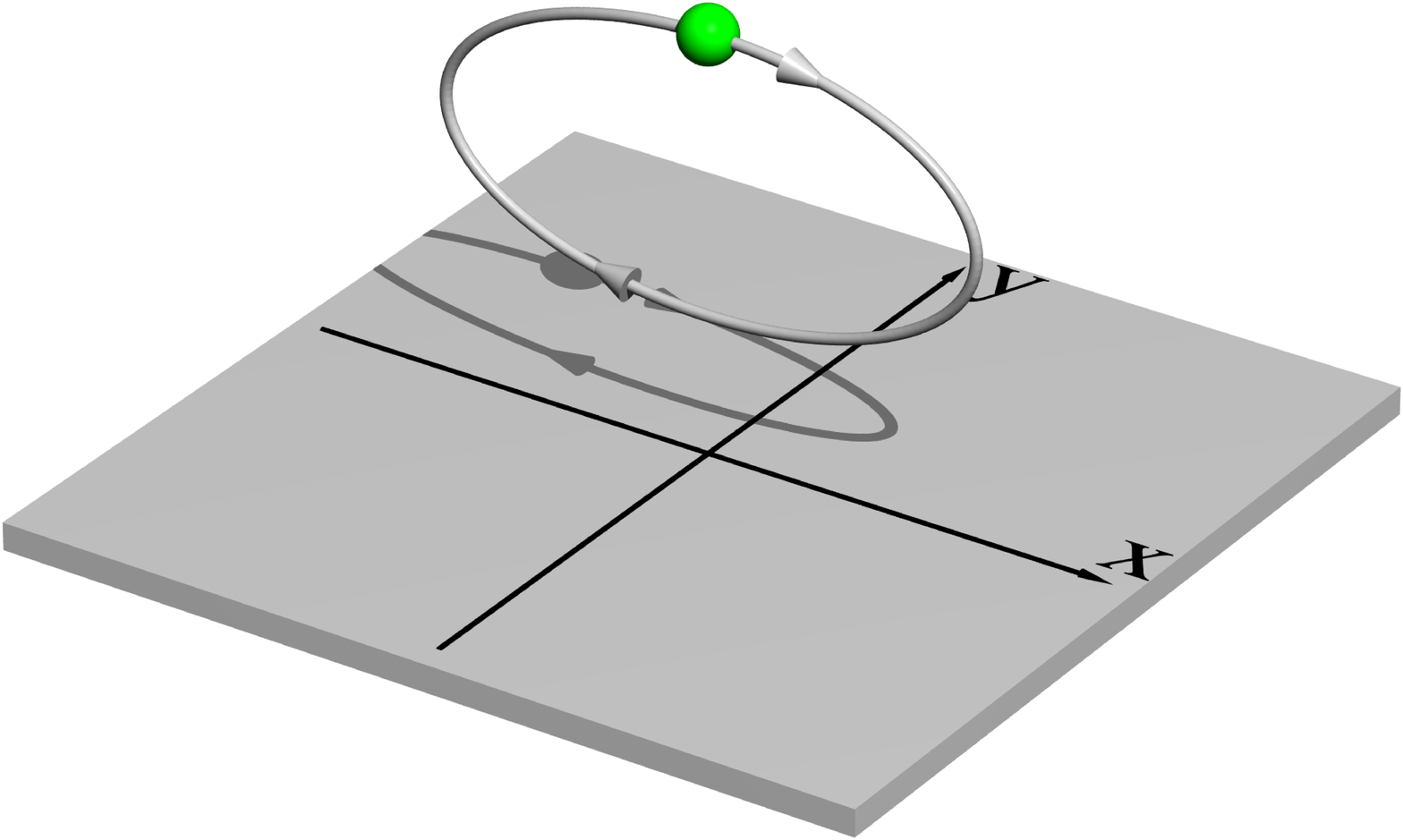}&
    c)\includegraphics[width=0.3\textwidth]{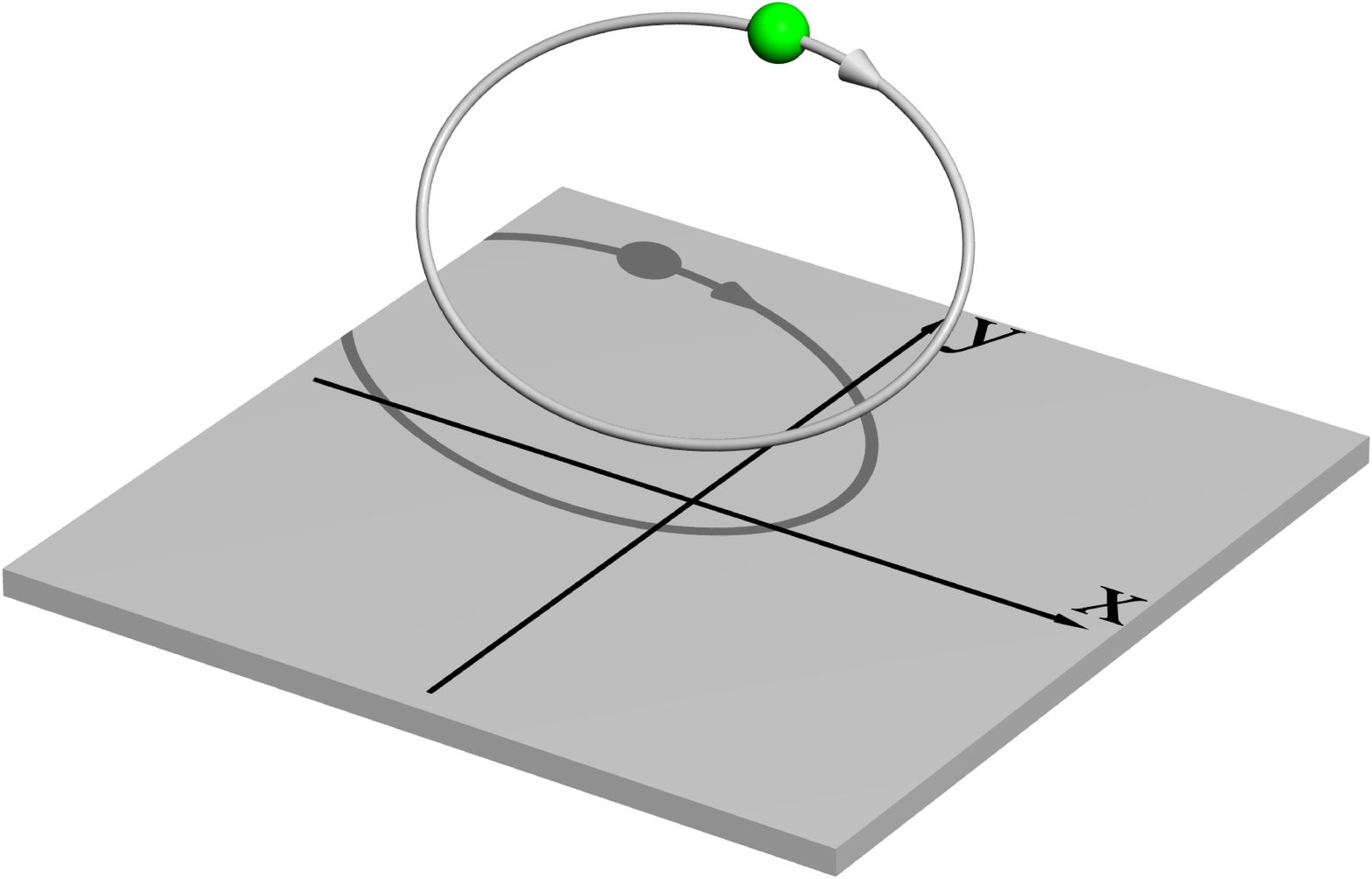}\\
    d)\includegraphics[width=0.3\textwidth]{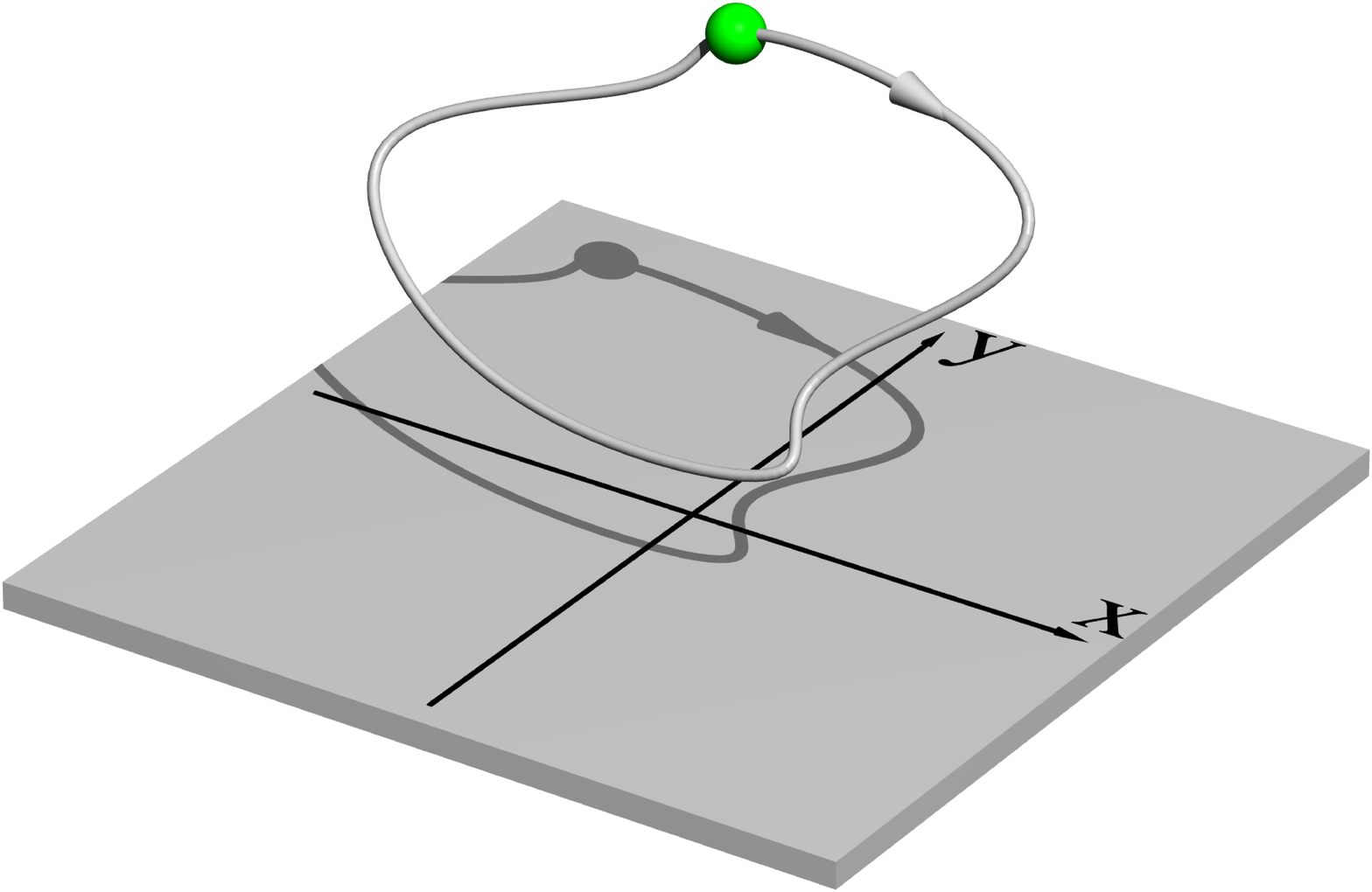}&
    e)\includegraphics[width=0.3\textwidth]{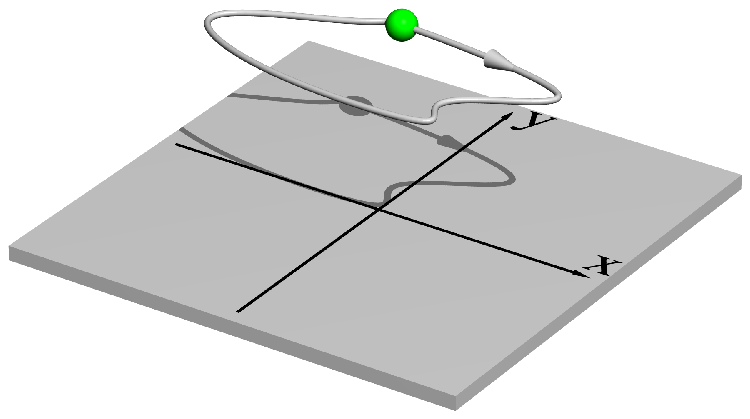}&
    f)\includegraphics[width=0.3\textwidth]{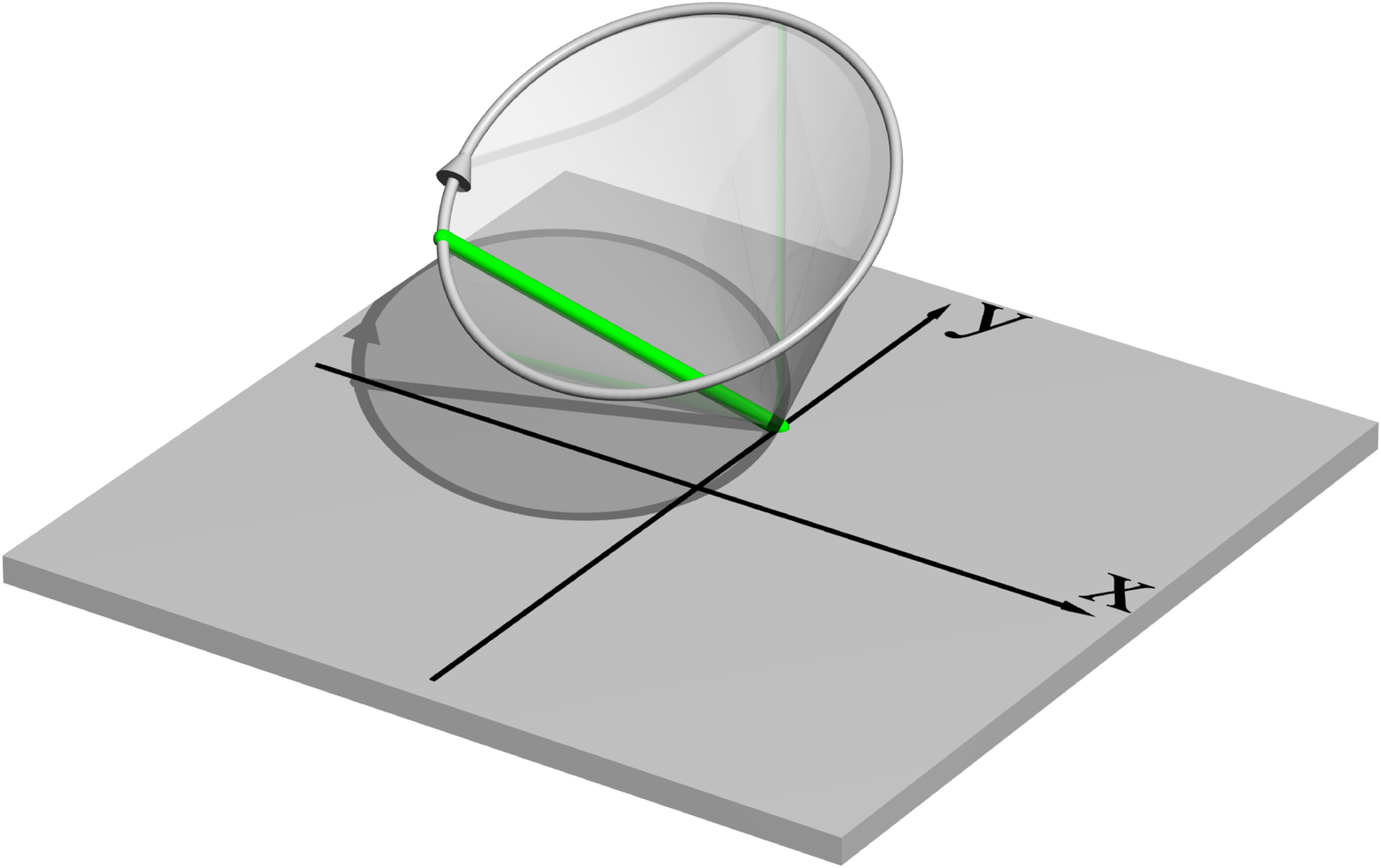}
  \end{tabular}
\end{center}
\caption{\label{fig:models}Models describing the ciliary beat as the motion of
  a small sphere along a closed trajectory: a) Linear, back-and-forth
  movement, which does not generate a directed flow. b) Planar elliptical
  trajectory. c) Tilted elliptical trajectory. d) General trajectory shape.
  e) General shape at a constant height. f) A model describing the cilium as a
  stiff slender rod.}
\end{figure*}

\subsubsection{Oscillating particle}
\label{sec:oscillating-particle}
We will start our discussion with a small particle, whose position is
oscillating periodically parallel to the $x$-axis (Fig.~\ref{fig:models}a). Of
course, this reciprocal motion will not generate any net (time-averaged) flow
\cite{Purcell1977}. However, this type of motion has been studied as a model
system for ciliary synchronization \cite{Kotar.Cicuta2010,Wollin.Stark2011}
and therefore its far-field is of interest. It should also be noted that even
if a single cilium performs reciprocal motion, an array of cilia with
metachronal coordination can still generate directed flow \cite{Lauga2011,Khaderi.Onck2012}.

Since the motion is symmetric with respect to the $x-z$ plane, the only
solutions that are theoretically possible are $D_x$, $Q_2$, $Q_3$ and $Q_5$. 

We parameterize the particle motion as
\begin{equation}
  \vec{R}(t)=\left( \begin{array}{c} B \sin \omega t\\0\\D
    \end{array}\right)
\end{equation}
so that the force acting on it is $\vec{F}=6\pi \eta a B \cos (\omega t)
\hat e_x$.
The resulting velocity field is 
\begin{multline}
  \vec{v}(\vec x,t)=\frac {3\omega a B D}{4}\bigl[ \sin(\omega t) \ (12
    D_x(\vec{x}) -6 D Q_3(\vec x)) \\- \sin (2\omega t) 3 B (Q_2(\vec
    x)-Q_5(\vec x)) \bigr]\;.
\end{multline}

\subsubsection{Planar beat}
\label{sec:planar-beat}
The next level of model is a planar ciliary beat. We model it with a small
particle moving along an elliptical trajectory in the $x-z$ plane
(Fig.~\ref{fig:models}b). Individually, planar cilia are rather inefficient
\cite{Downton.Stark2009}, but nearly planar beating patterns can be found in
respiratory epithelia, as well as in some microorganisms, such as \textit{Opalina}
\cite{Brennen.Winet1977}. Several models for ciliary synchronization study
planar beating patterns
\cite{Gueron.Blum1997,Guirao.Joanny2007,Lenz.Ryskin2006}.  Because of the
symmetry with respect to the $x-z$ plane, the solution will still consist of
the following terms alone: $D_x$, $Q_2$, $Q_3$ and $Q_5$.  The particle
trajectory is now parameterized as
\begin{equation}
  \vec{R}(t)=\left( \begin{array}{c} B \sin \omega t\\0\\C \cos \omega t + D
    \end{array}\right)
\end{equation}
and the force on the particle is $\vec{F}=6\pi \eta a (B \cos (\omega t)
\hat e_x-C \sin (\omega t) \hat e^z$. 

Based on the symmetry of the trajectory, which is invariant when we transform
$x\to -x$ and $t\to -t$ simultaneously, we conclude that all terms that are
even in $t$ have to be even in $x$. That reduces the possible terms to $D_x$
and $Q_3$. Conversely, the terms that are odd in time have to be odd in $x$,
which restricts them to $Q_2$ and $Q_5$.  The velocity field up to the
frequency $\omega$ (we omit higher harmonics, which exist up to the frequency
$3\omega$) reads
\begin{multline}
  \vec{v}(\vec x,t)=\frac {3\omega a}{4}\bigl[ 6BC D_x(\vec{x}) -6 BCD
  Q_3(\vec{x}) \\
+(12BD  D_x(\vec{x}) -6BD^2 Q_3(\vec{x})) \cos (\omega t)\\
+\frac 3 8 B^2 C ( Q_5(\vec{x}) -  Q_2(\vec{x}) ) \sin(\omega t) \bigr]\;.
\end{multline}

\subsubsection{Tilted ellipse}
\label{sec:tilted-ellipse}
More properties of a ciliary beat are captured by a model that describes it as
a sphere, circling on a tilted elliptical trajectory \cite{vilfan2006a}, as
shown in Fig.~\ref{fig:models}c. It is described as
\begin{equation}
  \vec{R}(t)=\left( \begin{array}{c} B \sin \omega t\\A \cos \omega t \\C \cos \omega t + D
    \end{array}\right)\;.
\end{equation}
The ($x\to -x$, $t\to -t$) symmetry is still present, so that the stationary
current can only consist of components $D_x$, $Q_1$, $Q_3$ and $Q_6$.  It is 
\begin{multline}
  \vec{\bar v}(\vec x)=\frac {3\omega a}{4}\bigl[ 6BC D_x(\vec{x}) -6 BCD
  Q_3(\vec{x}) -6 ABD Q_1(\vec{x})\bigr]\;.
\end{multline}

The major difference between the tilted ellipse and the vertical ellipse is
that the tilted one contains a $Q_1$ component, or a surface rotlet, as an
obvious consequence of the rotation around the $z$ axis. In the case of a
flat ellipse ($C=0$), this is the only stationary component. 

\subsubsection{General trajectory}
\label{sec:general-trajectory}
Let us finally discuss the stationary velocity field of a small particle
moving along an arbitrary periodic trajectory
(Fig.~\ref{fig:models}d). According to Eq.~(\ref{eq:velocity-gen}) the
velocity is
\begin{multline}
  \vec{\bar v}(\vec x)= \frac{9a}{T}\Bigl[D_x(\vec{x})  \oint Z dX + D_y(\vec{x})  \oint Z
  dY\\
-  \frac 12 Q_3(\vec{x})  \oint Z^2 dX - \frac 12 Q_4(\vec{x})  \oint Z^2 dY \\
+   \frac{Q_5(\vec{x})-Q_2(\vec{x})}{2}  \oint Z X dX 
+   \frac{Q_6(\vec{x})-Q_1(\vec{x})}{2}  \oint Z Y dX\\ 
+   \frac{Q_6(\vec{x})+Q_1(\vec{x})}{2}  \oint Z X dY
-   \frac{Q_5(\vec{x})+Q_2(\vec{x})}{2}  \oint Z Y dY \Bigr]
\label{eq:vgeneral}
\end{multline}
where $T$ denotes the period. The first two terms (through a coordinate
rotation they can be reduced to just $D_x$) generally dominate and they show
that the volume flow rate is proportional to the projection of the trajectory
onto the vertical plane in the direction of pumping.  An interesting special
case arises when the particle does not move vertically, $Z= {\rm const}$
(Fig.~\ref{fig:models}e). Then all integrals except those with $YdX$ and $XdY$
represent integration of a total differential along a closed path and
therefore vanish.  Equation (\ref{eq:vgeneral}) simplifies to
\begin{equation}
\label{eq:vrotating}
    \vec{\bar v}(\vec x)= \frac{9aZS}{T}Q_1(\vec{x})
\end{equation}
where $S$ is the area of the trajectory, $S= \oint X dY=-\oint Y
dX$. Regardless of the exact shape of the trajectory, a particle moving in the
horizontal plane only produces a surface rotlet in the order $r^{-3}$.

\subsection{Stiff slender rod}
\label{sec:stiff-slender-rod}

The next level of complexity is to describe the cilium as a thin stiff rod.
Although the hydrodynamic description in this model is more accurate than
replacing the cilium with a small particle, the motion is even more
restricted. At any time it is determined by two angles. There are different
ways to calculate the forces on the cilium. The simplest is the local drag or
resistive force theory (RFT) which assumes two constant friction coefficients
for motion parallel and perpendicular to the orientation of the rod
\cite{Lauga.Powers2009}.  It has the advantage that it allows analytical
calculation of the force distribution, while still giving good
results. Alternatively, one could use the slender body theory (SBT), which
includes hydrodynamic interactions for distances longer than a chosen cut-off
\cite{Johnson.Brokaw1979}. Even better accuracy can be achieved by describing
the cilium as a chain of spheres \cite{Gauger.Stark2009,Osterman.Vilfan2011},
a chain of regularized Stokeslets \cite{Ainley.Cortez2008} or with surface
boundary elements \cite{Smith.2009}.

Within the resistive force theory the local drag is described by a
tangential drag coefficient $C_T$ and a normal coefficient $C_N$ -- in
the case of a rod that is anchored with one end only the latter is
relevant. From the force density, we can calculate the far-field fluid flow
using Blake's tensor (\ref{eq:blake}) or its far field approximation
(\ref{eq:velocity-gen}).

A simple version of the slender rod model is obtained if the rod moves along
the mantle of a tilted cone, as proposed by Smith and coworkers
\cite{Smith.Gaffney2008}. The motion of a point on the cilium is then
described as (Fig.~\ref{fig:models}f)
\begin{equation}
\vec{R}=s \left( \begin{array}{ccc} \sin \psi \sin \omega t\\
\cos \theta \sin \psi \cos \omega t - \sin\theta \cos \psi\\
\sin \theta \sin \psi \cos \omega t +\cos \theta \cos \psi 
    \end{array}
  \right) + \frac {3L}{4}\left( \begin{array}{ccc}0 \\
\sin\theta \cos \psi\\
0
    \end{array}
  \right) \;.
\label{eq:rft-positions}
\end{equation}
with $s$ denoting the position parameter running from $0$ to $L$. 
We have placed the cone such that its center of weight, not the tip, lies on
the $z$ axis. We will soon see that this way we eliminated the terms
$Q_5$ and $Q_6$ and brought the flows to the normal form discussed in Section
\ref{sec:normalform}.  In RFT the force density on the cilium is simply the
transverse drag coefficient, multiplied by the local velocity
\begin{equation}
\vec{f}=C_N\, \omega s \left( \begin{array}{ccc} \sin \psi \cos \omega t\\
-\cos \theta \sin \psi \sin \omega t\\
-\sin \theta \sin \psi \sin \omega t 
    \end{array}
  \right)\;.
\label{eq:rft-forces}
\end{equation}
We obtain the far-field flow by inserting the force densities
(\ref{eq:rft-forces}) and positions (\ref{eq:rft-positions}) into
Eq.~(\ref{eq:velocity-gen}) and integrating over the length. The resulting
average flow is
\begin{multline}
  \vec{\bar{v}}(\vec{x})=\frac{C_N L^3 \omega}{4\pi \eta} \sin^2\psi
  \Bigl(\sin \theta 
    D_x
-\frac 3 4 L \cos^2 \theta \cos \psi Q_1  \\-\frac 3 4 L \sin \theta \cos
\theta \cos \psi Q_3\Bigr) \;.
\end{multline}
The first term is the well known surface Stokeslet term $D_x$, with an
amplitude proportional to $\sin\theta \sin^2\psi$ \cite{Smith.Gaffney2008}, or
the projection of the tip trajectory onto the $x-z$ plane. It is maximal for
$\psi=\arctan \sqrt 2$ and $\theta=\pi/2-\psi$.  The second term, $Q_1$, is a
surface rotlet and is maximal for $\theta=0$ (no tilt) and $\psi=\arctan \sqrt
2$. The third term, $Q_3$, a surface source-doublet, appears almost inevitably
along with the $D_x$ term.

An approximate value of the friction coefficient $C_N$ was given by Gueron and
Liron \cite{Gueron.Liron1992}
\begin{equation}
  C_N=\frac{8\pi \eta}{1+2 \ln (2q/a)}
\end{equation}
where $a$ is the radius of the rod and $q$ some characteristic length scale of
the order $\sqrt{La}$. A comparison with the Rotne-Prager approximation gives
a more precise estimate of $C_N=1.22\pi \eta$ for the ratio $L/a=20$
\cite{Vilfan.Babic2012}.

\begin{figure*}
  \begin{tabular}{cc}
    a)\includegraphics[width=0.4\textwidth]{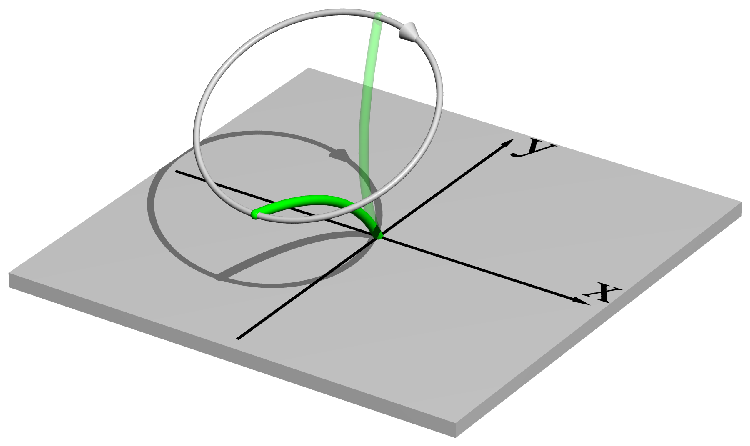}&
    b)\includegraphics[width=0.4\textwidth]{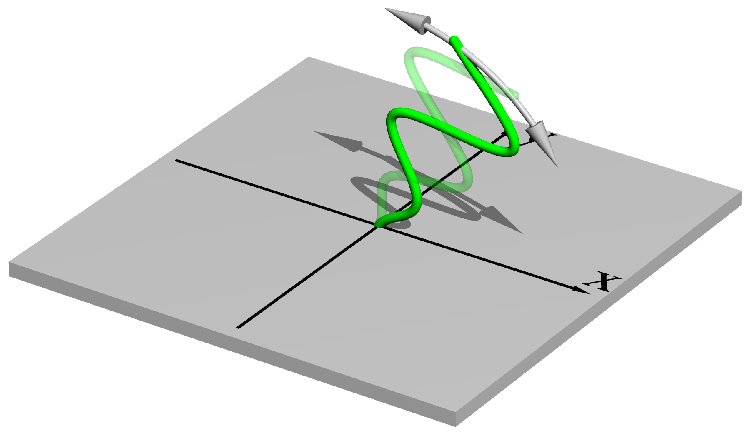}\\
    c)\includegraphics[width=0.4\textwidth]{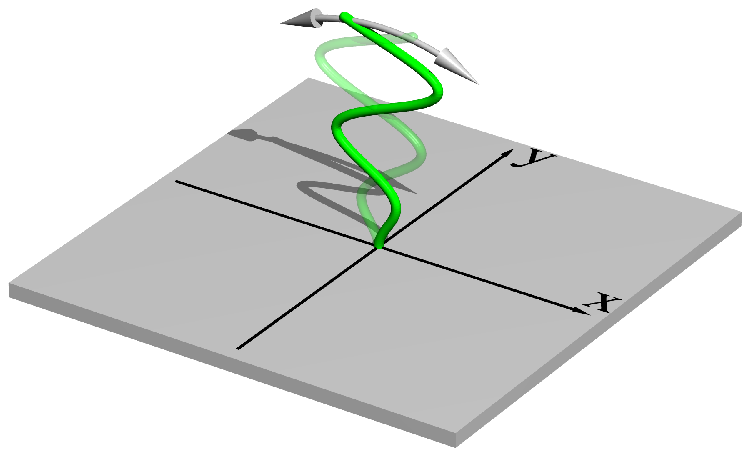}&
    d)\includegraphics[width=0.4\textwidth]{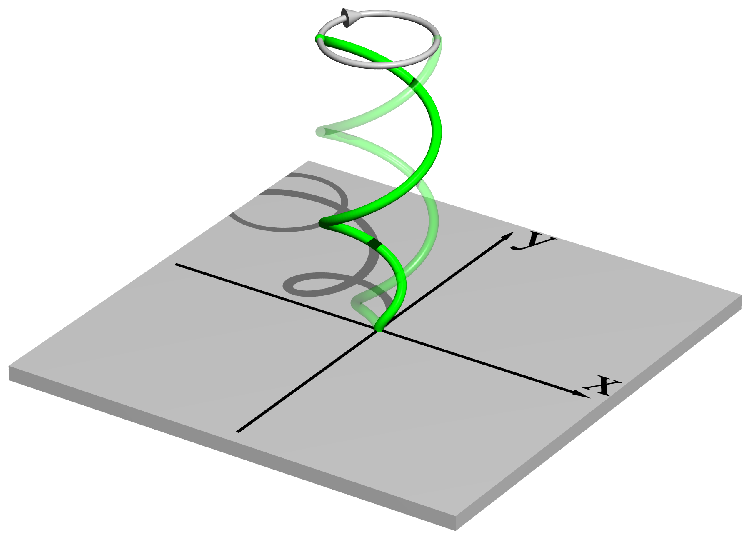}
  \end{tabular}
  \caption{Beating patterns according to their symmetry properties. In each
    figure the faint line shows the cilium half a period later. a) A pattern
    symmetric upon mirroring in $x$ direction and simultaneous time
    reversal. b) A planar beating pattern. c) A planar pattern that is
    symmetric with respect to $x\to -x$, $t\to t+T/2$. d) A pattern symmetric
    with respect to rotation around $z$ axis and simultaneous time shift. This
    corresponds to a cilium rotating around the $z$ axis.}
  \label{fig:symmetries}
\end{figure*}

\subsection{General}

In the following we will discuss the flows generated by an arbitrary periodic
beating pattern, based on its symmetry properties. Formally the problem could
be solved numerically within the resistive force theory \cite{Blake1974}, as
described in the previous section, although the motion would no longer be
purely transversal, so the tangential friction coefficient would need to be
included, too. Of course, any of the more accurate methods mentioned in the
previous section can be applied.

A common symmetry of the beating pattern includes mirroring in the $x$
direction and simultaneous time reversal ($x\to -x$, $t\to -t$),
Fig.~\ref{fig:symmetries}a. The patterns discussed in
Sections~\ref{sec:planar-beat},~\ref{sec:tilted-ellipse} and
\ref{sec:stiff-slender-rod} all contain this symmetry. But note that many
ciliary strokes found in nature do not. The typical recovery stroke during
which the cilium bends and sweeps along the surface is not
symmetric. Interestingly, the theoretically optimal solutions sometimes break
this symmetry spontaneously and sometimes not, depending on the allowed radius
of curvature \cite{Osterman.Vilfan2011}.

The symmetry property of the cilium trajectory has to be reflected in the
stationary flow component
\begin{equation}
\label{eq:symmetry}
  \vec{\bar v}(\tens{T}\vec{x})=s  \tens{T} \vec{\bar v}( \vec{x})
\end{equation}
with $s=-1$ and
\begin{equation}
  \tens{T}=\left( \begin{array}{ccc} -1 & 0 & 0 \\ 0& 1 & 0\\ 0& 0& 1
    \end{array}\right)\;.
\end{equation}
This condition is fulfilled by the modes $D_x$, $Q_1$, $Q_3$ and $Q_6$. Note
that this holds for the time-averaged flow (stationary component). The
oscillatory flow components can contain other modes.

Another interesting symmetric case is the general planar beating pattern
(Fig.~\ref{fig:symmetries}b). Then Eq.~(\ref{eq:symmetry}) holds with $s=1$
and
\begin{equation}
  \tens{T}=\left( \begin{array}{ccc} 1 & 0 & 0 \\ 0& -1 & 0\\ 0& 0& 1
    \end{array}\right)\;.
\end{equation}
It is fulfilled by modes $D_x$, $Q_2$, $Q_3$ and $Q_5$. An example of a planar
beating pattern is a cilium producing a waving pattern, similar to a planar
flagellum attached with one end to the surface. \textit{Opalina}
\cite{Brennen.Winet1977} produces this type of waves.

A further special case contains planar patterns that are additionally symmetric
with respect to the transformation $x\to -x$, $t\to t+T/2$
(Fig.~\ref{fig:symmetries}c). This imposes the additional symmetry with $s=1$
and
\begin{equation}
  \tens{T}=\left( \begin{array}{ccc} -1 & 0 & 0 \\ 0& 1 & 0\\ 0& 0& 1
    \end{array}\right)\;,
\end{equation}
which is only fulfilled by the models $Q_2$ and $Q_5$. An example is a planar
flagellum that beats in a way that is symmetric in $x$ direction, such that
the waves propagate vertically.

Let us finally look at a rotationally symmetric beating pattern
(Fig.~\ref{fig:symmetries}d). The symmetry transformation reads $x\to
\tens{T}x$, $t\to t + \Delta t$ with
\begin{equation}
  \tens{T}=\left( \begin{array}{ccc} \cos \varphi & \sin \varphi & 0 \\ 
      -\sin \varphi & \cos \varphi  & 0\\ 0& 0& 1
    \end{array}\right)
\end{equation}
and the two modes that fulfill this symmetry are $Q_1$ and $Q_2$. The first is
the obvious surface rotlet. Interestingly, a rotating flagellum can also
generate a vertical flow -- this is in contrast with a single circling
particle (\ref{eq:vrotating}). This type of flow is likely involved in the
otolith formation in the developing ear of the zebrafish embryo
\cite{Freund.Vermot2012,Wu.Vermot2011}.

\section{Conclusions}

Although the ciliary beating patterns can be very complex and is not yet well
understood from the mechanical perspective, their far-field properties up to
the order $r^{-3}$ can be described with only a few terms in a multipole
expansion. Of those, only the terms proportional to $r^{-2}$ are fundamental
-- all higher order terms can be represented as their spatial
derivatives. Depending on the symmetries present in the stroke pattern, the
number of terms describing the flow can often be further reduced. We have
derived the magnitudes of different terms for several models describing the
cilium as a small sphere, as well as for a rod-like cilium within the
framework of the resistive force theory. Of course, the multipole expansion is
not limited to any specific hydrodynamic approximation, but more accurate
descriptions require a numerical calculation of the amplitudes.

Although explicitly applied to cilia, our approach is equally suited for any
localized sources of flow in the proximity of a planar no-slip
boundary. Examples include an oscillating bubble
\cite{Marmottant.Hilgenfeldt2006}, a tumbling object
\cite{Sing.Alexander-Katz2010}, bacteria attached to the surface
\cite{Darnton.Berg2004} or swimming close to the surface
\cite{Mino.Clement2011} and many more.

Let us finally remark that our approach is not restricted to planar
boundaries. Exact solutions exist for flows due to point forces in the
vicinity of a sphere or inside a spherical cavity \cite{Maul.Kim1996}. Knowing
the fundamental solutions on a sphere (surface source and surface Stokeslet),
one can derive the higher order terms in a way similar to
Eqns.~(\ref{eq:d1deriv}--\ref{eq:d6deriv}). The derivation of the flows caused
by a single cilium of a spherical swimmer (e.g., \textit{Volvox}
\cite{Drescher.Tuval2010}) to the same order as shown here is then
straightforward.

\begin{acknowledgements}
  I would like to thank Frank Jülicher, Natan Osterman, Holger Stark, and
  Mojca Vilfan for discussions and Sascha Hilgenfeldt for comments on the
  manuscript.  This work was supported by the Slovenian Research Agency
  (Grants P1-0099 and J1-2209).
\end{acknowledgements}

%
% BibTeX users please use

\end{document}